\documentclass[11pt]{article}

\usepackage{graphicx}

\usepackage{amsfonts}
\usepackage{amsmath}

\def\xx{\mbox{\boldmath$x$}}
\def\xs{\mbox{\footnotesize\boldmath$x$}}

\def\T{^{\rm T}}
\def\av{^{\rm Av}}
\def\ot{\otimes}
\def\half{\tfrac{1}{2}}
\def\pmx{\begin{pmatrix}}
\def\emx{\end{pmatrix}}
\def\tsum{ \textstyle{\sum} }   
\newtheorem{lemma}{Lemma}
\newtheorem{corollary}{Corollary}

\title{Qubit Channels Which Require Four Inputs to Achieve Capacity: \\
    Implications for Additivity Conjectures}
\author{Masahito Hayashi\thanks{ERATO Quantum Computation and Information
Project, JST, Daini Hongo White Bldg. 201, 5-28-3, Hongo, Bunkyo-ku, Tokyo 113-0033, Japan}\ ,
Hiroshi Imai\addtocounter{footnote}{-1}\footnotemark\ 
\thanks{Department of Computer Science, University of Tokyo,
7-3-1, Hongo, Bunkyo-ku, Tokyo, 113-0033, Japan}\ ,
Keiji Matsumoto\addtocounter{footnote}{-2}\footnotemark\ 
\addtocounter{footnote}{1}\thanks{National Institute of Informatics (NII), 2-1-2 Hitotsubashi, Chiyoda-ku, Tokyo 101-8430, Japan}\ ,\\   
Mary Beth Ruskai\thanks{Department of Mathematics, Tufts University, Medford, Massachusetts  02155 USA }  
 \thanks{ partially supported  by
 the National Security Agency (NSA) and 
 Advanced Research and Development Activity (ARDA) under
Army Research Office (ARO) contract number 
     DAAD19-02-1-0065, and by the National Science
        Foundation under Grant  DMS-0314228.} 
 \,\,and Toshiyuki Shimono\addtocounter{footnote}{-3}\footnotemark
}

\footnotetext{ e-mail addresses: 
 ~ masahito@qci.jst.go.jp, imai@is.s.u-tokyo.ac.jp, keiji@nii.ac.jp,\\ MaryBeth.Ruskai@tufts.edu, shimono@is.s.u-tokyo.ac.jp\\ }
\date{\today}
\begin{document}

\maketitle

\begin{abstract}An example is given of a qubit quantum channel which requires
four inputs to maximize the Holevo capacity.   The example is one of a family 
of channels which are related  to 3-state channels.   The capacity
of the product channel is studied and numerical  evidence   presented 
which strongly suggests additivity.
The numerical evidence also supports a conjecture
about the concavity of output entropy as a function of entanglement 
parameters.   However,  an example is presented  which shows that 
 for some channels  this conjecture does not hold for all input
states.   A  numerical algorithm for finding the capacity and optimal
inputs is presented and its relation to a relative entropy optimization
discussed.
\end{abstract}

  \pagebreak

\section{Introduction}

The Holevo capacity $C(\Gamma)$ of a 1-qubit quantum channel $\Gamma$
is defined as the supremum over all possible ensembles of 1-qubit
density matrices $\rho_i$ and probability distribution $p_i$ 
of 
\[
S(\Gamma(\rho))-\sum_i p_i S(\Gamma(\rho_i))
\]
where $\rho=\sum_i p_i \rho_i$ is the average input and
$S(\sigma)=-{\rm Tr}(\sigma\log\sigma)$ denotes the von Neumann
entropy.
The Holevo capacity gives the maximum rate at which classical information 
can be transmitted
through the quantum channel \cite{holevo,schumacher-westmoreland}
using product inputs, but permitting entangled collective measurements.
It is a consequence of Carath\'eodory's Theorem and the convex
structure of this problem (as discussed in the next section) that
the above supremum can 
be replaced with the maximum over four input pairs of $(\rho_i,p_i)$.
(Davies \cite{davies} seems to have been the first to recognize
the relevance of Carath\'eodory's Theorem to problems  of this
type in quantum information theory; explicit application to
quantum capacity optimization appeared in \cite{fujiwara-nagaoka}).
 It was demonstrated in \cite{king-nathanson-ruskai}
that there exist qubit channels requiring three input states to
attain the maximum.   However, it was left open whether or not
there are 1-qubit
channels requiring four input states to achieve the maximum.  This paper
shows that such 4-input channels do exist by presenting an example.
The computation of this capacity is a nonlinear programming problem.
Unlike the classical channel capacity computation, this problem is
much harder, especially in a point that the classical case is the
maximization of a concave function while the quantum case is the
maximization of a function which is concave with respect to
probability variables, as in the classical case, and is convex with
respect to state variables.  As for algorithms to compute the capacity
by utilizing the special structure of the problem, \cite{nagaoka}
developed an alternating-type algorithm, by extending the well-known
Arimoto-Blahut algorithm for the classical channel capacity, and is
implemented in \cite{osawa-nagaoka} to check the additivity.  Use of
interior-point methods is suggested in \cite{imai-al}.  A method is 
presented in \cite{shor} for computing the capacity by combining
linear programming techniques, including column generation, with
non-linear optimization.    In this paper, we 
 present an approximation algorithm to compute the capacity of a
1-qubit channel; our  algorithm plays a key  role in finding a 4-state channel
numerically.
 
Although $C(\Gamma)$ plays an important role in quantum information
theory, it is not known whether or not using entangled inputs might
increase the capacity.   This is closely related to the question of the
additivity of  $C(\Gamma \otimes  \Gamma)$, which is now known 
\cite{pomeransky,MSW,shor2} to be equivalent to other conjectures including additivity 
of entanglement of formation.   In addition to being of interest
in their own right, 4-state channels are good candidates for testing
the additivity conjecture of the Holevo capacity for qubit channels.
We present numerical evidence for additivity which, in view of special
properties of the channels, gives extremely strong evidence for
additivity of both capacity and minimal output entropy for qubit channels.
Both results would follow from a new conjecture (which
appeared independently in \cite{DHS})  about concavity of
entropy as a function of entanglement parameters.    Using a 
different channel,
we show that this conjecture is false, at least in full generality.

The paper is organized as follows.    Basic background, definitions
and notation for convex analysis and qubit channels is presented
in Sections 2 and 3, respectively.    Numerical results for the
4-state channel and the
algorithm used to obtain them are described in Sections 4 and 5.
Some intuition about the properties of 3-state and 4-state channels 
is presented in Section 6 and shown to lead to additional examples
of 4-state channels.    In Section 7,  different views of the
capacity optimization are discussed and shown to be related
to a relative entropy optimization.    The additivity analysis and 
counterexample to the concavity conjecture are given in Section 8.
Throughout this paper, the base of the logarithm is 2.

\section{Convex Analysis} \label{sect:conv}

The function to be maximized in the Holevo capacity has a special form,
to which general convex analysis may be applied.  Based on \cite{rockafellar},
this section discusses the problem in this form.

Suppose $D$ is a $d$-dimensional bounded, closed convex set in ${\bf R}^d$,
and $f$ is a closed, concave function from $D$ to ${\bf R}$.
We are interested in the following infinite programming problem.
\begin{equation}
F=\sup_{\xs_i\in D, \, p_i} (f(\overline{\xx})-\sum_i p_i f(\xx_i))
\label{eqn:inf}
\end{equation}
where $\overline{\xx}=\sum_i p_i \xx_i$, $\sum_i p_i=1$, and
$p_i\ge0$.  This infinite mathematical programming problem
can be reduced to a finite mathematical programming
with $d+1$ pairs of $(x_i,p_i)$ as follows.

For such a closed, concave function $g$ over $D$, its {\it closure of
convex hull} function $\mbox{cl}~\mbox{conv}~g$ is the greatest convex
function majorized by $g$ (p.36, p.52 in \cite{rockafellar}).  In our
case, further using Carath\'eodory's Theorem (Theorem~17.1 in
\cite{rockafellar}), it is expressed as
\[
\mbox{cl}~\mbox{conv}~g(\xx)
=\min \bigg\{\, \sum_{i=1}^{d+1} p_i g(\xx_i) :
\xx=\sum_{i=1}^{d+1}p_i\xx_i,\ \sum_{i=1}^{d+1}p_i=1,\ 
\xx_i\in D,\  p_i\ge 0\ (i=1,\ldots,d+1)\,\bigg\}
\]
It is then seen that the problem (\ref{eqn:inf}) is reduced to the
following Fenchel-type problem (cf. Fenchel's duality theorem, section
31, \cite{rockafellar}).
\begin{equation}
\max_{\xs\in D}(f(\xx)-\mbox{cl}~\mbox{conv}~f(\xx)),
\label{eqn:fenchel}
\end{equation}
By virtue of nice properties of minimizing convex functions
 (e.g., Theorem~27.4  in \cite{rockafellar}), the optimality of a solution to this problem
is well-known:
\begin{lemma}
$\widetilde{\xx}$ is optimum in (\ref{eqn:fenchel}) if and only if there is
$\xi\in{\bf R}^d$ such that, for any $\xx\in D$,
\[
\xi\T(\xx-\widetilde{\xx})+\mbox{cl}~\mbox{conv}~f(\widetilde{\xx})
\le \mbox{cl}~\mbox{conv}~f(\xx)
\le f(\xx)
\le
\xi\T(\xx-\widetilde{\xx})+f(\widetilde{\xx}).
\]
Furthermore, when $f$ is strictly concave, there is a unique optimum
solution.
\end{lemma}
The above discussions can be summarized
in the form of problem (\ref{eqn:inf}) as follows:
\begin{corollary}
\label{cor:d+1}
In the infinite mathematical programming problem (\ref{eqn:inf}),
the supremum can be replaced with the maximum over $d+1$ pairs of
$(\xx_i,p_i)$.  
If there exist $d+1$ affinely
independent points $\xx_i$ $(i=1,\ldots,d+1)$ such that
a unique hyperplane passing through $(\xx_i,f(\xx_i))$ $(\xx_i\in D$,
$i=1,\ldots,d+1)$
in ${\bf R}^{d+1}$ is a supporting hyperplane to the convex
set $\{\,(\xx,y)\mid \xx\in D,\ \mbox{cl}~\mbox{conv}~f(\xx)\le y\le f(\xx)\,\}$
from below, and, for these $\xx_i$ $(i=1,\ldots,d+1)$, 
\[
\max\{\,f(\sum_{i=1}^{d+1} p_i \xx_i)-\sum_{i=1}^{d+1} p_i f(\xx_i)\mid
\sum_{i=1}^{d+1} p_i=1,\ p_i\ge0\,\}
\]
is attained with $p_i>0$ for all $i=1,\ldots,d+1$, then 
a set of $d+1$ pairs of $(\xx_i,p_i)$ is an optimum solution to (\ref{eqn:inf}).
\end{corollary}

\section{Set-up}
In the calculation of channel capacity for state on ${\bf C^d}$, the convex set 
$D$ is the set of density matrices, i.e., the set of  $d \times d$ positive
semi-definite matrices with trace $1$.   This is isomorphic
to a convex subset of ${\bf R}^{d^2-1}$.    A channel
$\Gamma(\rho)$ is described by a special type of linear map on the set of density matrices,
namely, one which is also  completely positive and trace-preserving.  

In the case of qubits, it is well-known that the set $D$  of density matrices
is isomorphic to the unit ball in ${\bf R}^3$ via the Bloch sphere representation.
We will use the notation $\rho(x,y,z)$ to denote  the density matrix
$\frac{1}{2}[ I + x \sigma_x + y \sigma_y +  z \sigma_z]$.   It was shown in
\cite{king-ruskai} that, up to specification of bases, a qubit channel can
be written in the form 
 \begin{eqnarray}  \label{eq:genform}
 \Gamma[\rho(x,y,z))] = \rho(\lambda_1 x + t_1 , \lambda_2 y + t_2 ,
     \lambda_3 z + t_3) .
\end{eqnarray}
 which gives an affine transformation on the Bloch sphere.  In fact, it
 maps the Bloch sphere
 $\{\,(x,y,z)\mid x^2+y^2+z^2\le 1\,\}$ to an ellipsoid with axes of
lengths $\lambda_1, \lambda_2, \lambda_3$ and center 
$t_1,t_2,t_3$.  Complete
 positivity poses additional constraints on the parameters
 $\{ \lambda_k, t_k \}$ which are given in  \cite{ruskai-szarek-werner}.

The strict  concavity of $S(\rho)$ 
implies that $S[\Gamma(\rho)]$ is also strictly concave for channels 
which are one-to-one.  In the case of qubits, this will hold {\em unless} the
channel maps the Bloch sphere into a one- or two-dimensional subset, which
can only happen when one of the parameters $\lambda_k = 0$.

\section{Numerical results}  \label{sect:num}

The theory in  Section~\ref{sect:conv}  can be used to calculate the capacity
with $f(\rho) = S[\Gamma(\rho)]$.   We are interested in qubit channels
with all $\lambda_k \neq 0$ so that  strict concavity  holds.
 Then the optimization problem as formulated
  in (\ref{eqn:fenchel})  has a unique solution.   However, 
 in the form  (\ref{eqn:inf}) as restricted in
Corollary~\ref{cor:d+1},  it may have multiple optimum solutions when
the hyperplane passes through more than $d+1$ such points.

Numerical optimization to compute the capacity of this channel
was initially performed by utilizing a mathematical programming 
package NUOPT \cite{nuopt} of Mathematical Systems Inc.   
These results, accurate to at most 7-8 significant figures, were
further refined by using them as starting points in a program
to find a critical point of the capacity by applying Newton's
method to the gradient.   The results are shown in Table~\ref{tab:4st}.
\begin{table}[here]
    $$\Gamma[\rho(x,y,z)] =  \rho(0.6x +0.21, 0.601y, 0.5z+0.495)   ~~\qquad ~~\text{capacity} =  0.3214851589 $$  \vskip-0.5cm

$$ S(\Gamma(\rho_i(x,y,z)))-\xi\T\Gamma(\rho_i) =  .9785055621 ~~~\forall ~ i \qquad 
   H[\Gamma(\rho_i),\Gamma(\rho_{\av})]  = 0.3214851589  ~~~\forall ~ i  $$ \vskip-0.5cm

     $$\begin{array}{ccrc}
 \text{probability}        &     \text{optimal  input } (x,y,z)         &     \phi  ~~~&   \theta         \\                  
 0.2322825705  & ( ~~0.2530759862,-0.0000000000,~~ 0.9674464043)   & ~~~0.127929 & 0     \\
 0.2133220819  &  (~~ 0.9783950999, ~0.0000000000, ~~0.2067438718) & 0.681275 & 0.0  \\
 0.2771976738  & (-0.4734087533, ~~0.8646461389,-0.1681404376)  &  0.869870 & ~2.071131  \\
 0.2771976738  &  (-0.4734087533,-0.8646461389,-0.1681404376)  &    0.869870 & -2.071131    \\
 \text{average}     &  ( ~~0.0050428099, ~~0.0000000000, ~~0.1756076944) &   \end{array}  $$
     \flushright {$\phi, \theta$ denote the angular coordinates of the optimal inputs  .}  \vskip-0.2cm

 $$  \begin{array}{ccc}
 \text{probability}       &     \text{optimal  output } (x,y,z)      & S[\Gamma(\rho)]     \\
 0.2322825705 &~~( 0.1728455917, 0.0000000000, 0.9787232022)   & 0.0300135405  \\
0.2133220819  &    ( 0.6080370599, 0.0000000000, 0.5983719359)  & 0.3786915585   \\
0.2771976738  & (-0.2630452520, 0.5196523295, 0.4109297812)  &  0.5935800377   \\
 0.2771976738   &~~ (-0.2630452520,-0.5196523295, 0.4109297812) & 0.5935800377  \\
 \text{average}  & ( 0.0240256859, 0.0000000000, 0.5828038472)  & 0.7383180644 \end{array}  $$

 \caption{Data for 4-state channel }
\label{tab:4st}
\end{table}

\begin{figure}[p]
 \begin{center}
 \includegraphics*[width=8cm,height=8cm,keepaspectratio=true]{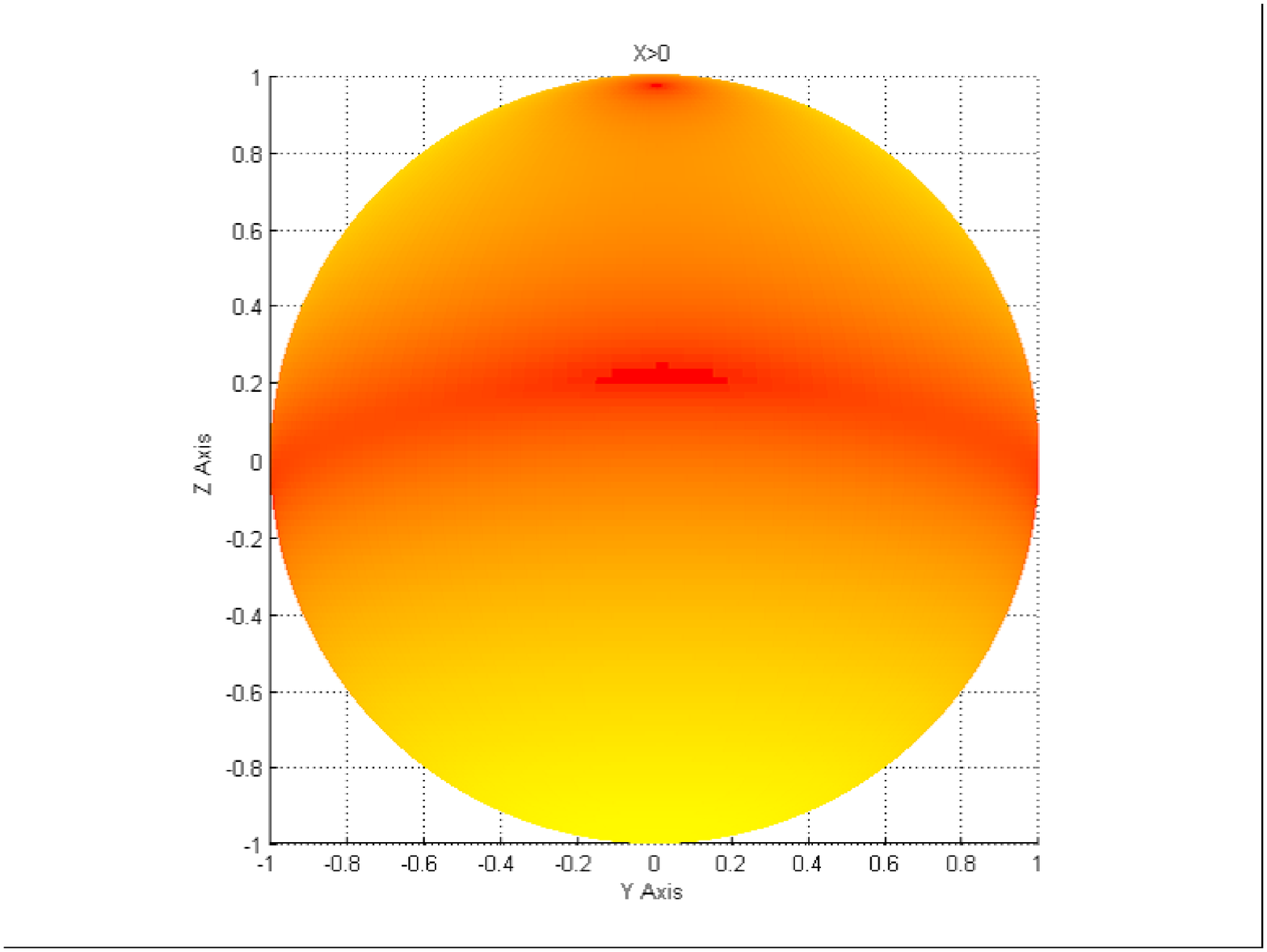}
\includegraphics*[width=8cm,height=8cm,keepaspectratio=true]{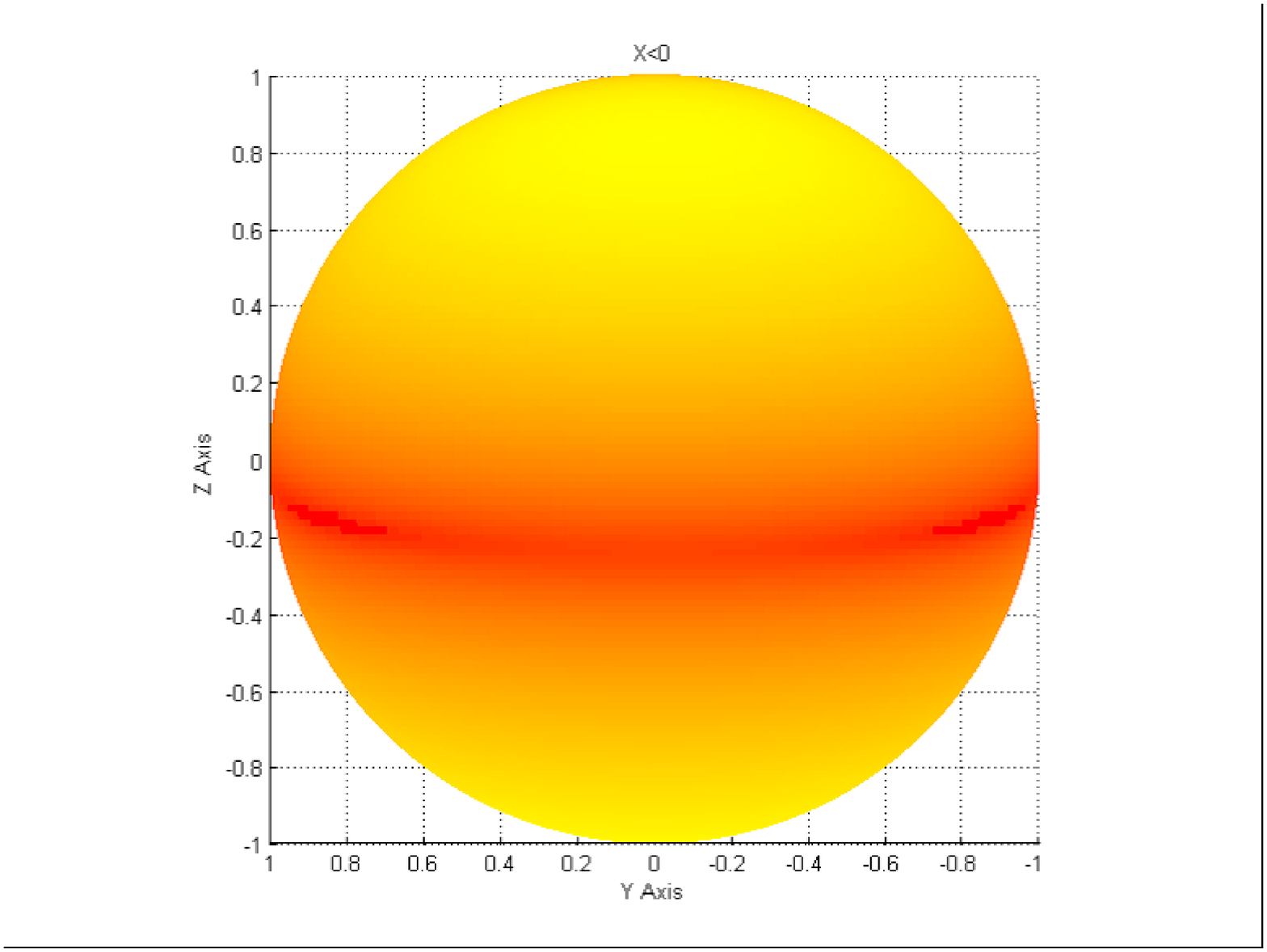}

inputs $ (\rho(x,y,z)$ on the two hemispheres of the Bloch sphere.   ~~left:  $x > 0$~~~~right:  $x < 0$ \\
  \end{center}
    \begin{center}
\includegraphics*[width=8cm,height=8cm,keepaspectratio=true]{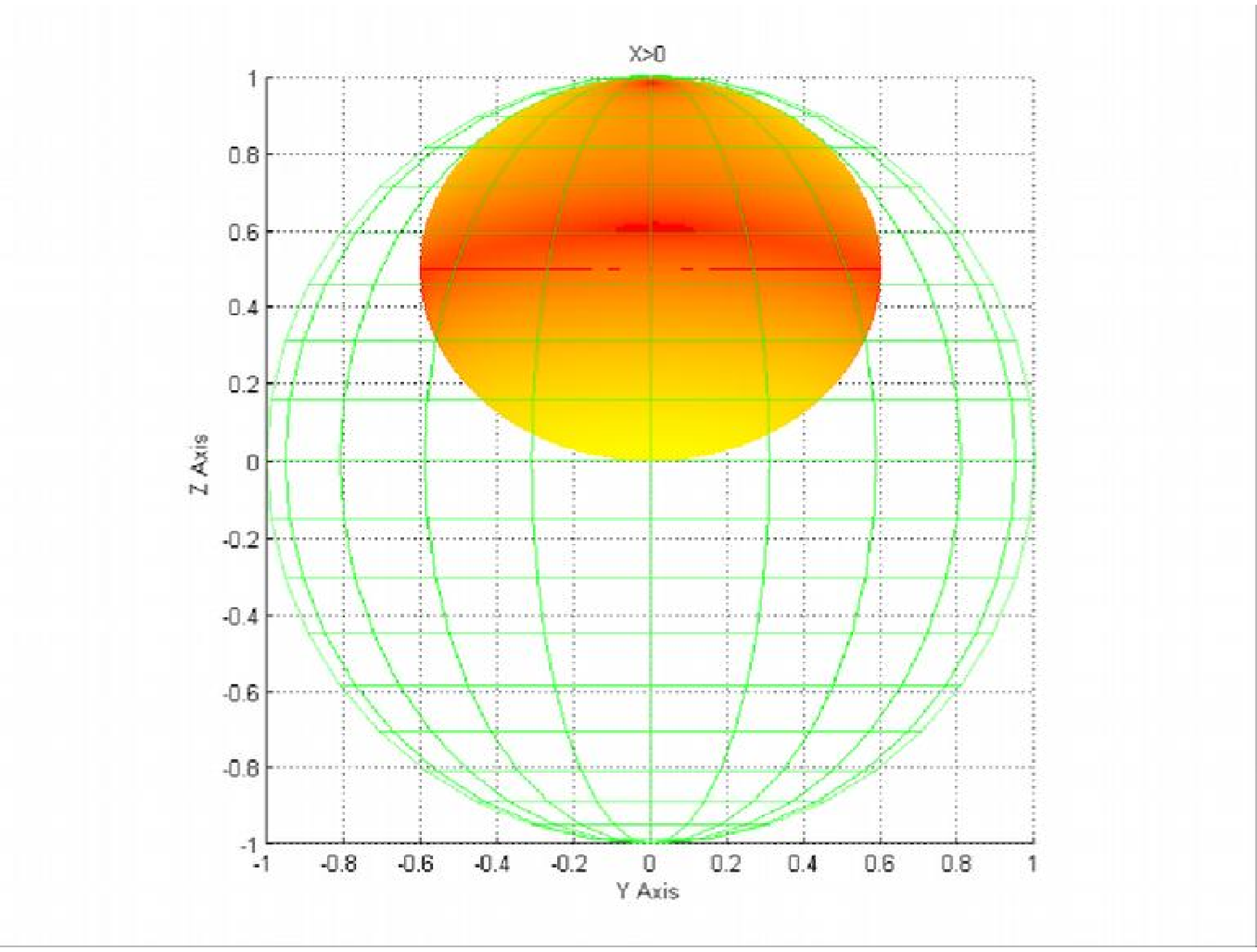}
\includegraphics*[width=8cm,height=8cm,keepaspectratio=true]{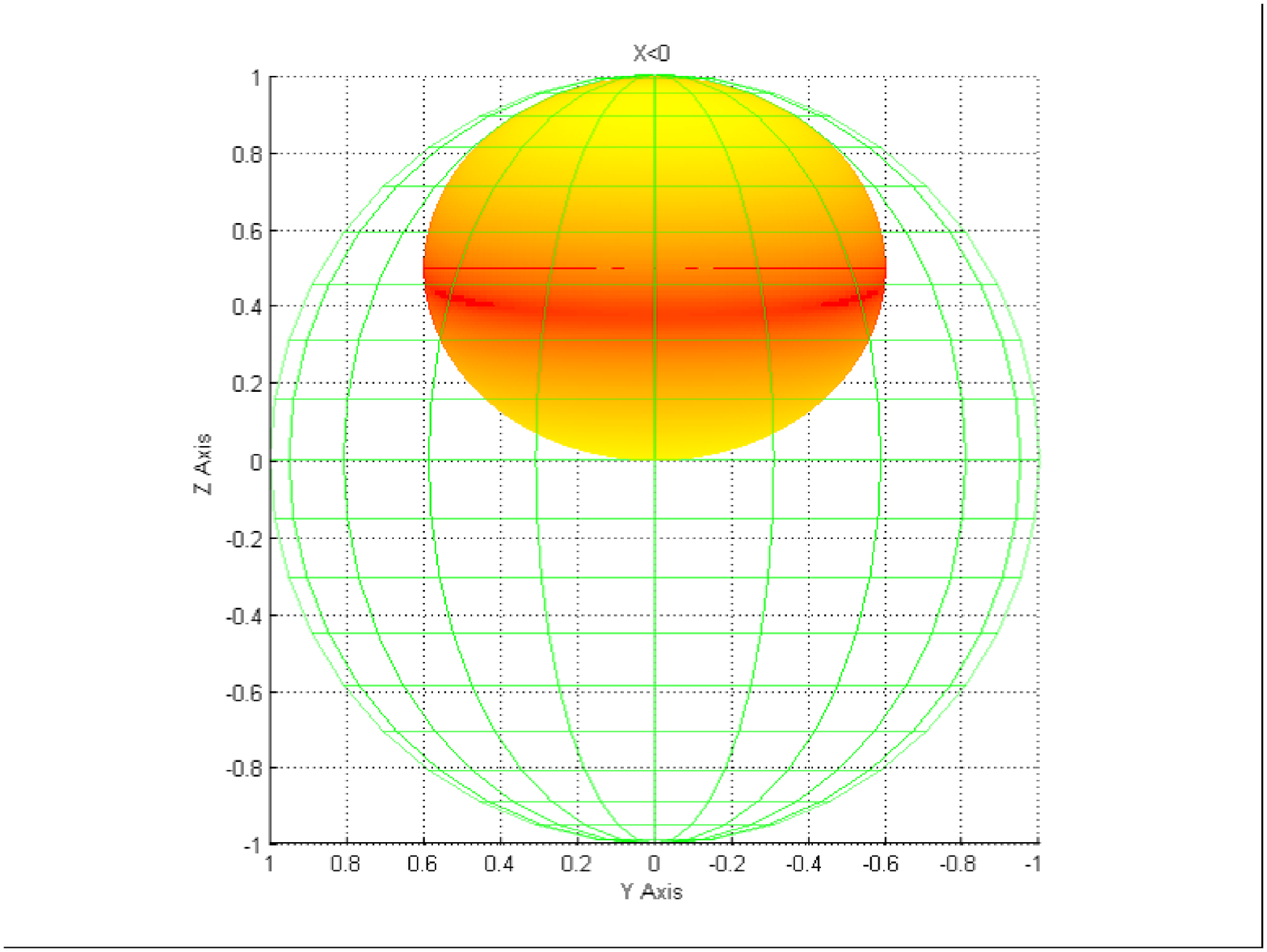}

output states $\Gamma(\rho(x,y,z)) $  on the image ellipsoid. ~~left:  $x > 0$~~~~right:  $x < 0$
   \end{center}
 
 \vskip0.5cm
      \begin{center}
    \includegraphics*[height=1.5cm,keepaspectratio=true]{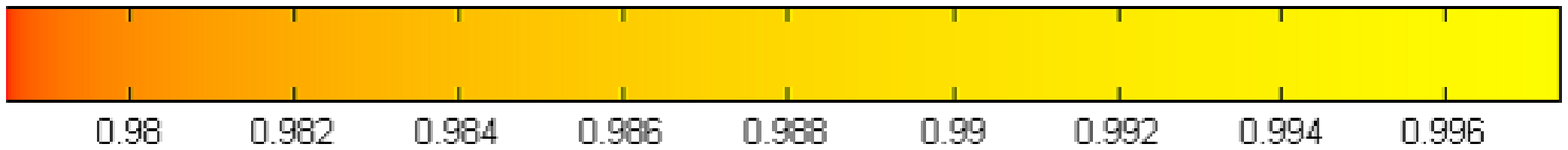}
\newline     Scale for interpretation $F(x,y,z) = S(\Gamma[\rho(x,y,z)])-\xi\T\Gamma[(\rho(x,y,z)])$
     \vskip0.5cm
~~~~  \includegraphics*[height=2.0cm,keepaspectratio=true]{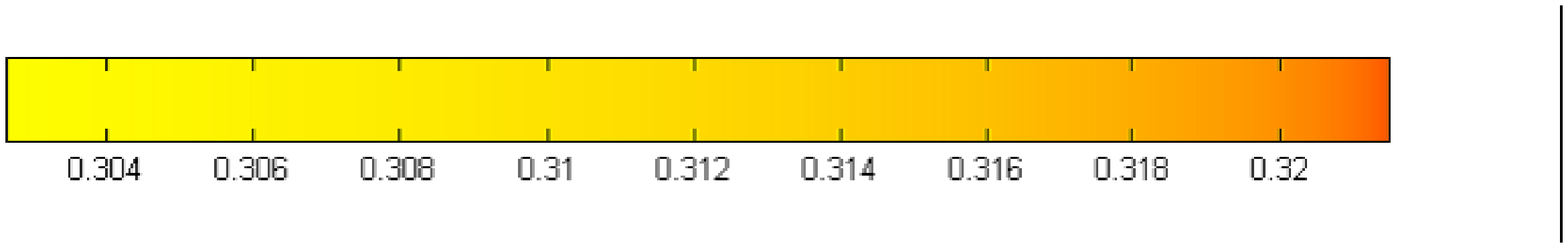}
 \newline    Scale for interpretation as  $H[\Gamma(\omega), \Gamma( \rho^4_{\av})] $   \end{center}
      \vskip0.5cm
  \caption{Depiction of $F(x,y,z) = S(\Gamma[\rho(x,y,z)])-\xi\T\Gamma[(\rho(x,y,z)])$ and
relative entropy  $H[\Gamma(\omega), \Gamma( \rho^4_{\av})]   = 1.299989 - F(x,y,z) $
with respect to optimal average output
   in terms of color (or grey scale) on the boundary of the Bloch sphere and its image.}
   \label{fig:four}
  \end{figure}

To verify  that these results give a true 4-state optimum, the function
  $S(\Gamma(\rho(x,y,z)))-\xi^{\T}\Gamma(\rho)$ was computed and plotted
 with $\xi = (-0.0396622022, 0 , -0.9621071440)$. 
These results are shown in Figure~\ref{fig:four}
and confirm the condition that the hyperplane  $(\xi,-1)\cdot (x,y,z,w) = -0.9785055621 $
passes through the four points  $((x_i,y_i,z_i ,S(\Gamma[\rho(x_i,y_i,z_i)]) )$
and the condition that the hyperplane lies below the surface   $(x ,y ,z ,S(\Gamma[\rho(x,y,z)]) $ in ${\bf R}^{4}$. 
(The components $\xi_x,\xi_y,\xi_z$ of $\xi$ are obtained by solving the four simultaneous equations
$\xi^T \cdot \Gamma(\rho_k) +\xi_0=S(\Gamma[\rho(x_i,y_i,z_i)]) \;(k=1,2,3,4)$
for the variables $(\xi_x,\xi_y,\xi_z,\xi_0)$. )
  As discussed in Section~\ref{sect:relent} this is equivalent to a   relative
entropy optimization.

In addition, the optimal three-state capacity
was also computed and shown to be $ < 0.321461$ which is strictly less
than the 4-state capacity of  0.321485.     Details for the 3-state capacity
can be found in Table~\ref{tab:3st} (Section \ref{sect:heur}).
As an optimization problem, the capacity has other local maxima 
in addition to the  3-state and 4-state results discussed above.
For example, there  are several 2-state optima, but these have lower
capacity and are not relevant to the work presented here.

\section{Approximation Algorithm to Compute the Holevo Capacity}

To find the  4-state channel given above, the following
approximation algorithm was repeatedly applied with various parameters.
This approximation algorithm is almost sufficient to compute the Holevo
capacity of a 1-qubit channel in practice.

Recall that the problem (\ref{eqn:inf}) is an infinite mathematical
programming problem.  As far as all $\xx_i\in D$ are considered, this
infinite set may be regarded as fixed, leaving only  $p_i$ as variables.
 The objective function is concave with respect to $p_i$, which
is quite nice to solve, although the problem is still an infinite
one.

\begin{figure}[h]
  \includegraphics[width=8cm]{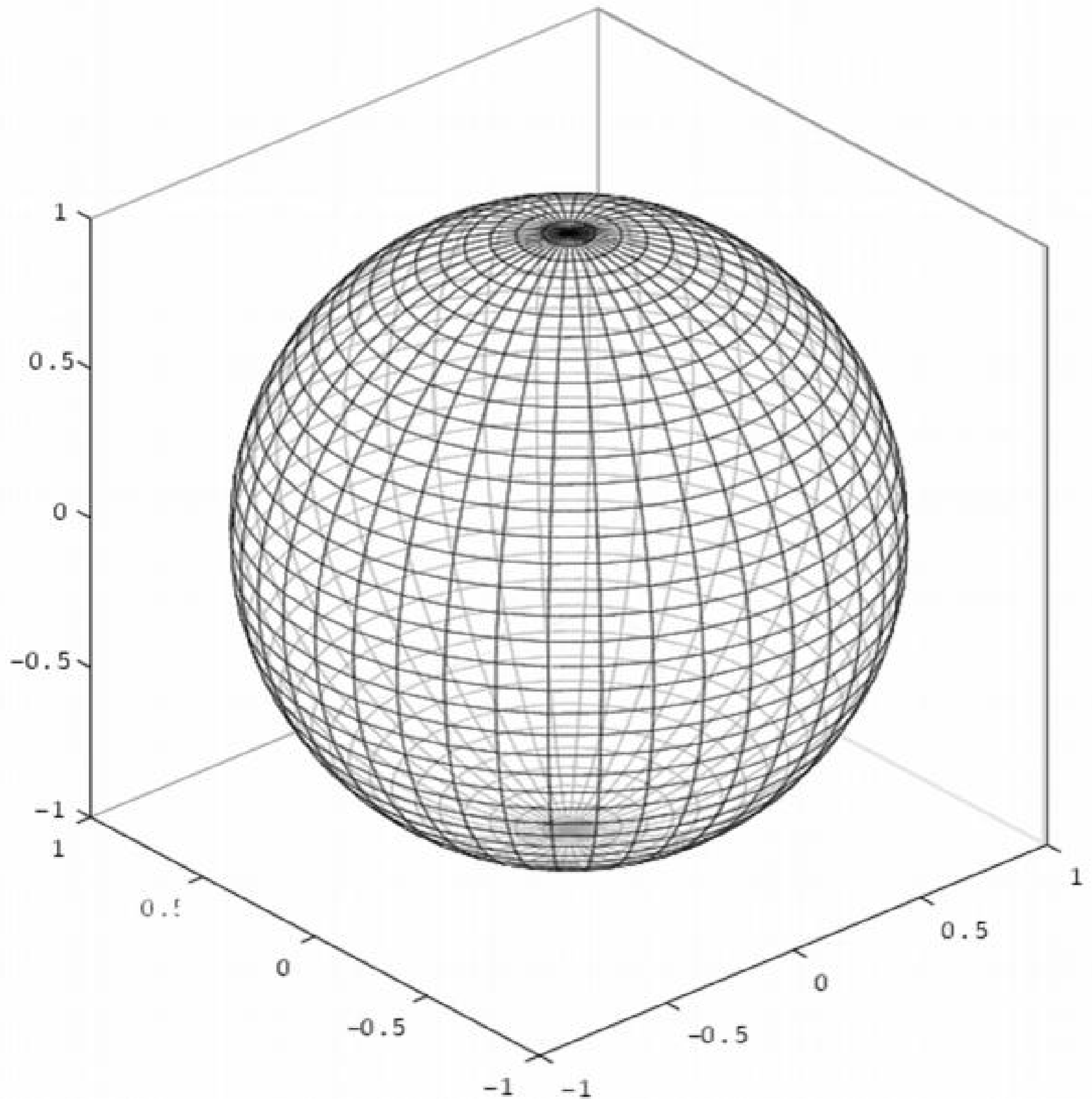}
  \includegraphics[scale=0.45]{figure2b.eps}
\vspace{0.6cm}
\caption{
(left)
A polyhedral approximation of the sphere for $k=40$. It has $k^2-k+2=1562$ points.
(right)
Approximation values by $(k^2-k+2)$-point mesh.  The horizontal axis
is a log plot of $k$, and the vertical axis is a log plot of the difference
to the optimum value in bit.  A line $y=0.05/x^2$ is drawn for reference.}
\label{fig:mesh}
\end{figure}

For a 1-qubit channel, owing to the concavity of the von Neumann
entropy, in the formulation (\ref{eqn:inf}), $\xx$ can be restricted to a
pure state, i.e., $x^2+y^2+z^2=1$ in terms of the Bloch sphere.
The sphere is two-dimensional, and the convex hull of 
a square mesh of $k(k+1)$ points
$(\sin(\theta_j)\cos(\mu_l),\sin(\theta_j)\sin(\mu_l),\cos(\theta_j))$
with
$\theta_j=j\pi/k$, $\mu_l=2l\pi/k$ $(j=0,...,k$; $l=0,\ldots,k-1)$
is quite a good polyhedral approximation.  For $j=0,k$ and any $l$, points become
$(0,0,1)$ and $(0,0,-1)$, and the total number of points is $k^2-k+2$
(See Fig.\ref{fig:mesh}, left).
Then, considering the problem of type (\ref{eqn:inf}) for these
$k^2-k+2$ points with constraints $\sum_{i=1}^{k^2-k+2} p_i=0$,
$p_i\ge0$, the maximum to this $(k^2-k+2)$-dimensional concave maximization
problem gives a close lower bound to the real maximum of the original
problem.

Interior-point methods can be applied to this high-dimensional
concave maximization programming problem (e.g., \cite{potra-ye}).
Computational results from NUOPT  are shown in Fig.\ref{fig:mesh}, right,
from which this approximation approach provides  values sufficiently
close to the Holevo capacity in practice.

\section{Heuristic construction of a 4-state channel}  \label{sect:heur}
 
The existence of four state channels of the type found above
can be understood as emerging from small deformations of
3-state channels with a high level of symmetry.
As noted above, a channel   of the form (\ref{eq:genform})
maps the Bloch sphere  to an ellipsoid with axes of
lengths $\lambda_1, \lambda_2, \lambda_3$ and center 
$t_1,t_2,t_3$.   When $t_1 = t_2 = t_3$, the ellipsoid is centered
at the original and the capacity is achieved with a pair of orthogonal
inputs which map to the endpoints of the longest axis of the 
ellipsoid.      However, when some $t_k$ are non-zero, this no 
longer holds and it can even
happen that the capacity is achieved with a pair of orthogonal
inputs which map to the endpoints of the shortest axis   
(as for the example $ \Gamma[\rho(x,y,z))] = \rho(0.55x, 0.55y, 0.5z +0.5)$.)
By finding parameters which balance these situations,
3-state channels were  constructed in \cite{king-nathanson-ruskai} .

One of the 3-state channels in \cite{king-nathanson-ruskai}  is
\begin{eqnarray} \label{eq:stretch}
\Gamma(\rho(x,y,z))=\rho(0.6 x, 0.6 y, 0.5 z +0.5)
\end{eqnarray}
which has rotational symmetry about the z-axis 
of the Bloch sphere.  This allows one to analyze the problem in 
two-dimensional plane, but with the limitation that at most a 3-state
channel can be found.   Although the analysis of this channel was
performed in the $x$-$z$ plane, one could,  instead,  choose the optimal 
inputs to lie any plane
containing the $z$-axis, e.g., the $y$-$z$ plane.  Moreover,  if one
 replaces the two inputs $(\pm 0.93681, \, 0,\,  -0.34984)$,
each with probability 0.29885,  by any three or more states with $z = -0.34984$ 
which also average to  $(0, 0, -0.34984)$ the capacity is unchanged.
However, only three inputs are actually necessary to achieve this capacity.

To find a true 4-state channel, the symmetry must be lowered 
so that the full 3-dimensional geometry of the Bloch sphere is
required.    The channel (\ref{eq:stretch}) was obtained as a convex
combination of an amplitude damping channel with
$ \lambda_1 = \lambda_2 = \frac{1}{\sqrt{2}} \approx 0.707$
and a shifted depolarizing channel with $ \lambda_1 = \lambda_2 = 0.5$.
Thus, once could expect to make minor changes to 
$  \lambda_1 $ and/or $\lambda_2$ without violating the
 CP condition \cite{fujiwara-nagaoka,king-ruskai, ruskai-szarek-werner} of
 $ (\lambda_1 \pm  \lambda_2)^2 \leq (1 \pm  \lambda_3)^2 + t_3^2
   = \frac{ 9 \pm 1}{4} $   for channels with $t_1 = t_2 = 0$.
Letting
   $  \lambda_1  = 0.6,   \lambda_2 = 0.601$ gives a
channel with reflection symmetry across the $x$-$z$ and $y$-$z$ planes.
Its capacity will require three input states  which lie
in the $y$-$z$ plane as shown in Table ~\ref{tab:3st}.
We now wish to further reduce the symmetry by shifting the
ellipsoid.  To do so, one must first decrease $\lambda_3$ or $t_3$.
We consider the channel
\begin{eqnarray}  \label{eq:chan2}
 \Gamma[\rho(x,y,z))] = \rho(0.6 x  , 0. 601y , 0.5 z + 0.495)
\end{eqnarray}
which is still CP and requires three input states 
which lie in the $y$-$z$ plane as shown in Table ~\ref{tab:3st}.

 \begin{table} [p] 
    $\Gamma(\rho(x,y,z))=\rho(0.6 x, 0.6 y, 0.5 z +0.5)$ \hskip2cm  $C(\Gamma) = 0.324990$ 
 \vskip-0.5cm $$     \begin{array}{ccc}
    \text{probability}  &   \text{inputs}   &   \text{outputs} \\
   0.402338  & ( 0.000000,  0.000000,  1.000000)   & ( 0.000000,  0.000000,  1.000000)   \\
 0.298830  & ( 0.936786   \cos \vartheta ,  0.936786   \sin \vartheta  , -0.349902 )   & ( 0.562072  \cos \vartheta ,  0.562072   \sin \vartheta  ,  0.3250492)  \\
 0.298830  & ( -0.936786  \cos \vartheta , - 0.936786   \sin \vartheta  , -0.349902)   & ( -0.562072  \cos \vartheta ,  -0.562072   \sin \vartheta  , 0.325049 )  \\
\text{average}       & ( 0.000000,  0.000000,  0.193215 )   & ( 0.000000,  0.000000,  0.596608)
 \end{array} $$   \vskip0.1cm  
 The ellipsoid is symmetric about the $z$-axis so that  the optimal inputs can be chosen to lie in
 any plane containing the $z$-axis.   For $\vartheta = \pi/2$, $x = 0$ and  the optimal inputs lie in 
 the $y$-$z$ plane;
  for $\vartheta = 0$, $y = 0$ and   the optimal inputs lie in the $x$-$z$ plane.  
     \\ ~~ \\ ~~ \\
       $\Gamma(\rho(x,y,z))=\rho(0.6 x, 0.601 y, 0.5 z +0.5)$ \hskip2cm  $C(\Gamma) = 0.325555 $ 
 \vskip-0.2cm $$    \begin{array}{ccc}
    \text{probability}  &   \text{inputs}   &   \text{outputs} \\
     0.380692  & ( 0.000000,  0.000000,  1.000000)   & ( 0.000000,  0.000000,  1.000000)   \\
 0.309653  & ( 0.000000 ,  0.952435 , -0.304740 )   & ( 0.000000,  0.572413 ,  0.347630)  \\
 0.309653  & ( 0.000000,    -0.952435  , -0.304740 )   & ( 0.000000, -0.572414,  0.347630)  \\
\text{average}       & ( 0.000000,  0.000000,  0.191964 )   & ( 0.000000,  0.000000,  0.595982)
\end{array}  $$ \vskip0.1cm  
   The longest axis of the ellipsoid is parallel to the $y$-axis, and optimal inputs 
  lie in the $y$-$z$ plane.
   \\ ~~ \\ ~~ \\
         $\Gamma(\rho(x,y,z))=\rho(0.6 x, 0.601 y, 0.5 z +0.495)$ \hskip2cm  $C(\Gamma) =   0.320535 $ 
 \vskip-0.2cm $$    \begin{array}{ccc}
    \text{probability}  &   \text{inputs}   &   \text{outputs} \\0.146660 & ( 0.000000,  0.000000,  1.000000)   & ( 0.000000,  0.000000,  0.995000)  \\
 0.426670  & ( 0.000000,  0.999687,  0.025034)   & ( 0.000000,  0.600811 ,  0.507517 )  \\
 0.426670  & ( 0.000000, -0.999687,  0.025034 )   & ( 0.000000, -0.600811 ,  0.507517 )  \\
\text{average}       & ( 0.000000,  0.000000,  0.168022 )   & ( 0.000000,  0.000000,  0.579011 )
\end{array}  $$ \vskip0.1cm  
           The longest axis of the ellipsoid is parallel to the $y$-axis, and optimal inputs 
  lie in the $y$-$z$ plane.
       \\ ~~ \\ ~~ \\
      $\Gamma(\rho(x,y,z))=\rho(0.6 x+ 0.021, 0.601 y, 0.5 z +0.495)$ \hskip2cm  
          $C_3(\Gamma) =  0.3214609877$ 
  \vskip-0.5cm  $$    \begin{array}{cccc}
    \text{probability}  &   \text{inputs}   &   \text{outputs} & S[\Gamma(\rho)]  \\ 
   0.213290   &  ( 0.252867, 0.000000, 0.9675017) & ( 0.172720, 0.000000, 0.978751) & 0.029992  \\
 0.366051 & ( 0.978544, 0.000000, 0.206036 ) & ( 0.608127, 0.000000, 0.598018)  & 0.379029\\
 0.420657 & ~ (-0.967649 , 0.000000,-0.252299) ~& ~  (-0.559590, 0.000000, 0.368851) & 0.645884 ~\\
 \text{average}    &    ( 0.005083, 0.000000, 0.1756493) & ( 0.024050, 0.000000, 0.582825)
& 0.738297  \end{array}  $$   \vskip0.1cm  
 A shift in the $x$-direction offsets the slightly greater length
  parallel to the $y$-axis so that the {\em restricted} 3-state optimization  inputs lie
  in the $x$-$z$ plane.  However, the 3-state capacity is less than that for the unrestricted 
    problem which requires four input states.  
\vskip0.5cm
     \caption{Optimal 3-state ensembles  for various channels}
\label{tab:3st}
\end{table}

We now shift the channel in the x-direction and study
\begin{eqnarray}  \label{eq:chan3}
 \Gamma[\rho(x,y,z))] = \rho(0.6 x +0.21 , 0. 601y , 0.5 z + 0.495)
\end{eqnarray}
The CP condition [12,13] for a channel of the form
\begin{eqnarray}  \label{eq:chan3cp}
 \Gamma[\rho(x,y,z))] = \rho(0.6 x + t_1 , 0. 601y , 0.5 z + 0.495)
\end{eqnarray}
reduces to $\det( I - R^T R) \geq 0$ where
$ R =  \begin{pmatrix} \frac{t_1}{\sqrt{ (1.995)( 0.005)}} &  
   \frac{1.201}{\sqrt{ (1.995)( 1.005)}} \\  & \\
   \frac{-0.001}{\sqrt{ (0.995)( 0.005)}}&
\frac{t_1}{\sqrt{(0.995)(1.005)}}  \end{pmatrix}  . $
This gives the quartic inequality
 $ 0.2805326349- 101.0098436\,t_1^2+ 100.2531329\,t_1^4 \geq 0 $ 
which  holds for  $|t_1| \leq 0.05277$. .

Although small enough to satisfy the CP condition, a shift of $t_1= 0.021$
is sufficient  to return the (restricted)
3-state optimum to the $x$-$z$ plane across which the image has
reflection symmetry.    In fact, the inputs 
$\rho(x,y,z)$ and $\rho(x,-y,z)$  have the same output entropy.  
Moreover, replacing {\em all} inputs $\rho_i(x,y,z)$ by $\rho_i(x,-y,z)$
leaves the capacity unchanged.  Therefore, 
 {\em either} all optimal inputs lie in the $x$-$z$ plane 
{\em or} the set of optimal inputs contains pairs
 of the form  $\rho(x,\pm y,z)$ with the same probability.  
 (This follows easily from a small modification of the   convexity argument 
 in \cite{king-nathanson-ruskai}. )  Let 
 \begin{eqnarray}   \label{eq:chi}
 \chi[ \pi_1, \rho_1 , \pi_2,  \rho_2, \pi_3,  \rho_3 ] 
   =  S\big( \tsum_i \pi_i \rho_i \big) - \sum_i     \pi_i S(\rho_i) .        
 \end{eqnarray}     
For simplicity,  assume  that  $y_1 = y_2 = 0$, but $y_3 \neq 0$.
Let $\pi_4 = \pi_3$ and $\rho_4 =  \rho(x_3,-y_3,z_3)$. Then
 \begin{eqnarray*}   
\lefteqn{ \chi\big[ \pi_1, \rho_1 , \pi_2,  \rho_2 , \half \pi_3,  \rho_3,   
      \half \pi_3,  \rho_4 \big]    = 
         } \\ & ~ & \half  \chi\big[ \pi_1, \rho_1 , \pi_2,  \rho_2,  \pi_3,  \rho_3 \big]  +
        \half   \chi\big[ \pi_1, \rho_1 , \pi_2,  \rho_2,  \pi_4,  \rho_4 \big]
  +   S(\widetilde{\rho}) -  \half S\big( \sum_{i=1,2,3}  \pi_i \rho_i \big)  
          -  \half S\big(  \sum_{i=1,2,4} \pi_i \rho_i \big) \\
          & =  &  \chi[ \pi_1, \rho_1 , \pi_2,  \rho_2, \pi_3,  \rho_3 ] +
             S(\widetilde{\rho}) -  \half S\big( \sum_{i=1,2,3}  \, \pi_i \rho_i \big)  
          -  \half S\big(  \sum_{i=1,2,4} \, \pi_i \rho_i \big) \\
            & > &  \chi[ \pi_1, \rho_1 , \pi_2,  \rho_2, \pi_3,  \rho_3 ] =
                  \chi[ \pi_1, \rho_1 , \pi_2,  \rho_2, \pi_3,  \rho_4 ]
  \end{eqnarray*}   
  where $\widetilde{\rho} = \pi_1 \rho_1 + \pi_2 \rho_2 +  \half \pi_3 \rho_3 +  \half \pi_4 \rho_4 
  =  \displaystyle{\half  \sum_{i=1,2,4} \pi_i \rho_i +  \half \sum_{i=1,2,4} \pi_i \rho_i}$.  
   The  strict inequality then
  follows from the strict concavity of $S(\rho)$. 
  
  To see why one might expect a 4-state optimum with one pair of inputs with 
  $\pm y$ and two with $y_i =0 $,   consider the effect of replacing a
  state of the form $(A,0,B)$ by a pair of the form $( A^{\prime}, \pm(a+b), B^{\prime})$ 
  with $(A^{\prime})^2 = A^2 - a^2 $, $(B^{\prime})^2 = B^2 - b^2 $.    
  Recall that increasing the length of an output state decreases the entropy 
  and, hence, increases the capacity; moreover this effect is greatest  when the 
  changes to the output   are orthogonal to the level sets $x^2 + y^2 + z^2 = const$
  of entropy.     For our channel,  increasing $y_i$ with $a,b$ having the opposite sign
  of $A,B$ will increase the contribution of $-S[\Gamma(\rho_i)]$ to the capacity.
  But one must also consider the competing effect of these changes on
  $S[\Gamma(\rho_{\av})]$ for which the net result depends on the
  geometry of the image.   Since $\Gamma(\rho_{\av})$ is near
  $(0,0,0.5)$, changes in $x,y$ will have little effect on the entropy.
  However, decreasing $z$ will move the average closer to $\half I$ in
  a direction near that of greatest increase in entropy.
  Comparison of the results in Tables 1 and 2 shows
  results consistent with this analysis, but more complex due to the various
  competing effects.   Roughly speaking, the input at 
  $ (-0.967649 , 0.000000,-0.252299)$ with entropy  
  $S[\Gamma(\rho_3)] = 0.645884$  splits into the pair
  of inputs $  (-0.473409, \pm 0.864646,-0.168140)  $ with output
  entropy $S[\Gamma(\rho_i)] = 0.593580$.
  However, decreasing $|z_i|$ increases $z_i$ in this case; this
  is offset by changing  $\pi_i = 0.4207  $
  to a pair with $p_i = 0.2772$ increasing the net weight to $0.5544$ for the states
  with negative $z_i$.    But the new  
  outputs still have  higher entropy than those from  inputs with positive $z_i$,
    The net result  is that 
  the average outputs of  $ ( 0.024050, 0.000000, 0.582825) $ and 
  $ ( 0.024026, 0.000000, 0.582804)   $  are very close for the
  3-state and 4-state optima,  and the increase from
   3-state to 4-state capacity  is only about $1.5  \times 10^{-5}$.
        
     \begin{figure}[p] 
\begin{center}   

  
   \includegraphics*[width=10cm,height=10cm,keepaspectratio=true]{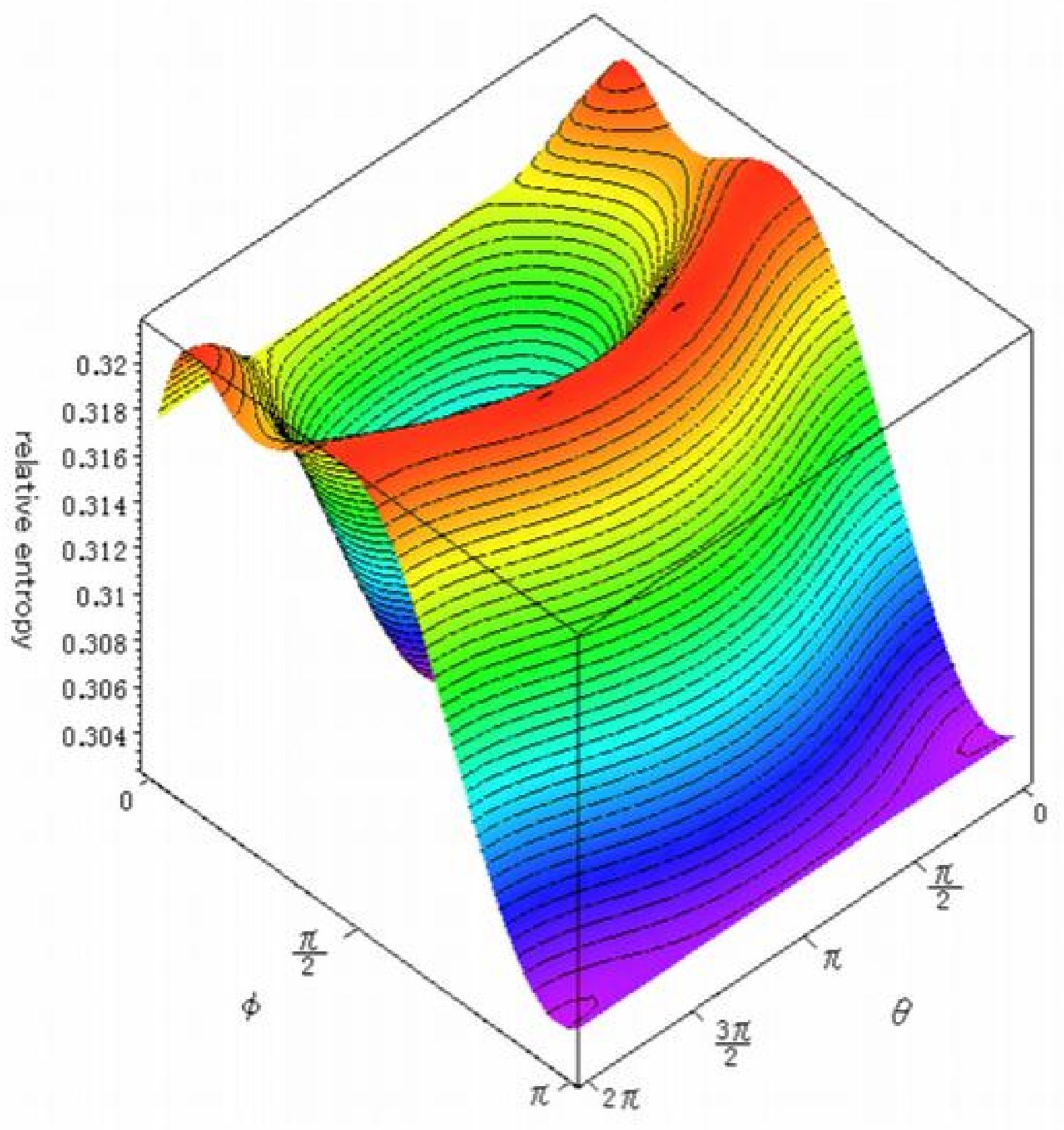}   

 Plot of $H[\Gamma(\omega(\cos \theta \sin \phi, \sin \theta \sin \phi ,\cos \phi)], \Gamma( \rho^4_{\av})] $.

 \includegraphics*[width=8cm,height=8cm,keepaspectratio=true]{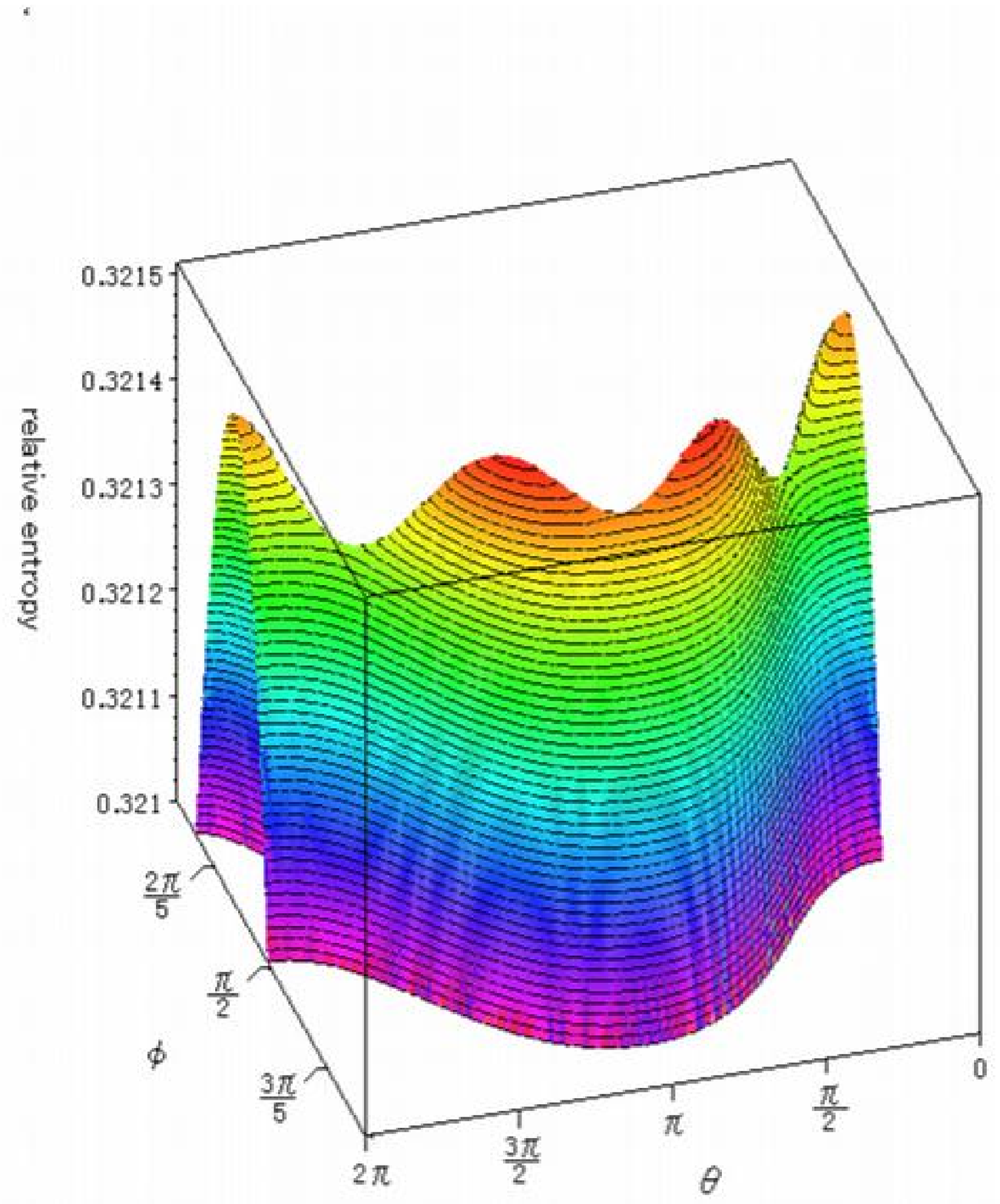}   
 
    Detailed view of the dark ``ridge''
    near $\phi = \frac{\pi}{2}$ showing 3 distinct maxima and saddle points.

    \label{fig:relent3}
    \caption{Plots of relative entropy of output states with respect to the optimal average output  as 
    a function of a pair of angles defining  pure input states on the surface of the  Bloch sphere.   
     The edges $\theta = 0$ and $\theta = 2 \pi$ meet on the sphere, so that
the figures show two halves of the two maxima with $y = 0$, one
near the north pole and one on the ridge. }
   \end{center}
  \end{figure}

The 4-state channel found in Section~\ref{sect:num} is not unique.  For example, the channel
$\Gamma(\rho(x,y,z))=  \linebreak \rho(0.8 x + 0.22, 0.8015 y, 0.75 z + 0.245)$ also requires
4-states to optimize capacity.    In view of the discussion above
it is reasonable to expect that one can
find a family of 4-state channels which have the form
$\Gamma(\rho(x,y,z))= \rho(\lambda_1 x + \epsilon_1, (\lambda_1 + \epsilon_2) y, \lambda_3 z + t_3)$
with  $\epsilon_k$ suitable small constants,
$ \lambda_3  + t_3 = 1 - \epsilon_3$, and $ \lambda_1 > \lambda_3$  chosen so that
$\Gamma(\rho(x,y,z))= \rho(\lambda_1 x ,  \lambda_1  y, \lambda_3 z + t_3)$ is
close to a 3-state channel.   

In the class of channels above, one always has $t_2 = 0$, which raises the question of
whether or not there exist    4-state channels exist with all $t_k$ all non-zero.
Therefore, maps  of the form
 $\Gamma(\rho(x,y,z))=  \rho(0.6 x + 0.021, 0.601 y+ t_2, 0.5 z + 0.495)$
were considered with $t_2 \neq 0$.  With $t_2 < 0.48$ such maps are completely 
positive and the channel with 
with $t_2 = 0.00005$ was shown to require four inputs to achieve
capacity.

   \section{Equivalence to a relative entropy optimization} \label{sect:relent}
   
   Reformulation of the capacity optimization in the dual form (\ref{eqn:fenchel})
   was also used by Audenaert and Braunstein \cite{AB} and by Shirokov \cite{shirokov}
   to obtain theoretical results and plays an important role in Shor's proof \cite{shor2}
   of equivalence of additivity questions.      
   The implication that the optimal outputs for the capacity then define a supporting 
   hyperplane  for the output entropy function $S[\Gamma(\rho )]$   can also be
   reformulated in terms of relative entropy.

The relative entropy is defined as
 $H(\omega,\rho) \equiv  {\rm Tr} \;\omega (\log \omega - \log \rho)$.
It then follows that
\begin{eqnarray}
      S(\rho)  - \sum_i \pi_i S(\rho_i) = \sum_i  \pi_i H(\rho_i, \rho)
\end{eqnarray}
and
\begin{eqnarray}  \label{eq:rel.sup}
       C(\Gamma) = \sup_{\rho} \sup_{\pi_i,\rho_i} \Big\{ \sum_i  \pi_i H[\Gamma(\rho_i), \Gamma( \rho)]  :  
          \tsum_i \pi_i \rho_i = \rho, \pi_i > 0, \tsum_i \pi_i = 1  \Big\} .
\end{eqnarray}
Moreover, for any fixed $\rho$, and any $\pi_i, \rho_i$
\begin{eqnarray}  \label{eq:relentsqz}
   \sum_i  \pi_i H[\Gamma(\rho_i), \Gamma( \rho)]  ~  \leq ~
       C(\Gamma) ~  \leq ~ \sup_{\omega}  H[\Gamma(\omega), \Gamma( \rho)]  .   
       \end{eqnarray}
In fact, it was shown in \cite{ohya-petz-watanabe} and \cite{schumacher-westmoreland:relent} that
\begin{eqnarray} \label{eq:maxmin}
        C(\Gamma)=  \inf_{\rho}  \sup_{\omega}  H[\Gamma(\omega), \Gamma( \rho)]  
    \end{eqnarray}
from which it follows  that when $\rho_{\av}$ is the {\em optimal average input},
$C(\Gamma)= H[\Gamma(\rho_i), \Gamma( \rho_{\av})]$ for all $i$.   Thus, a necessary
condition that an ensemble ${\mathcal E} = \{ \pi, \rho_i \}$ achieve the capacity
is that all outputs $\Gamma(\rho_i)$ are ``equidistant'' from the average output
$\Gamma \big( \sum_i  \pi_i \rho_i \big)$ in the sense that
$H[\Gamma(\rho_i), \Gamma( \rho_{\av})]$ is independent of $i$.

\begin{figure}[p]
  \vskip2cm
     \begin{center}
  \includegraphics*[width=8cm,height=8cm,keepaspectratio=true]{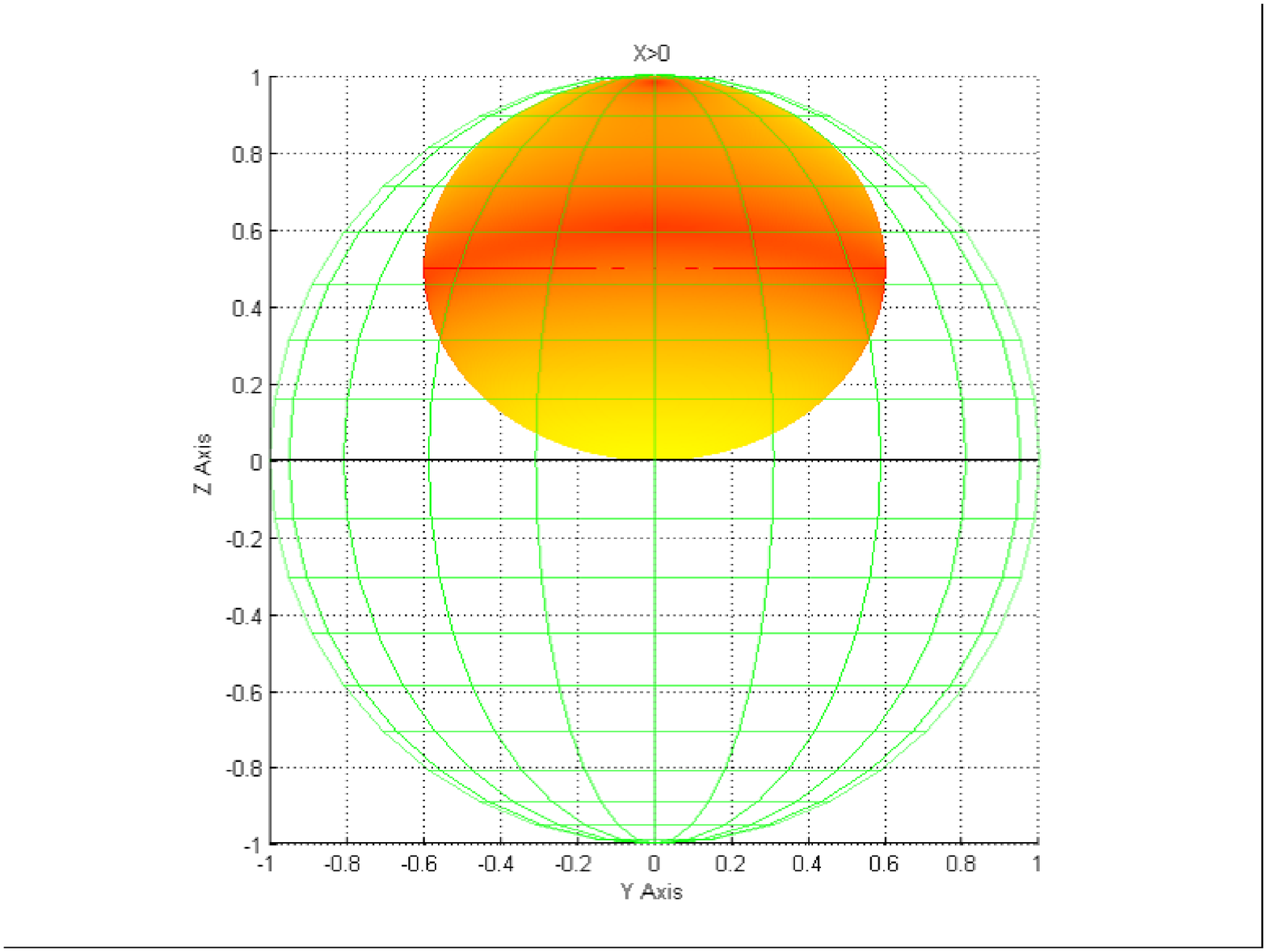}
 \includegraphics*[width=8cm,height=8cm,keepaspectratio=true]{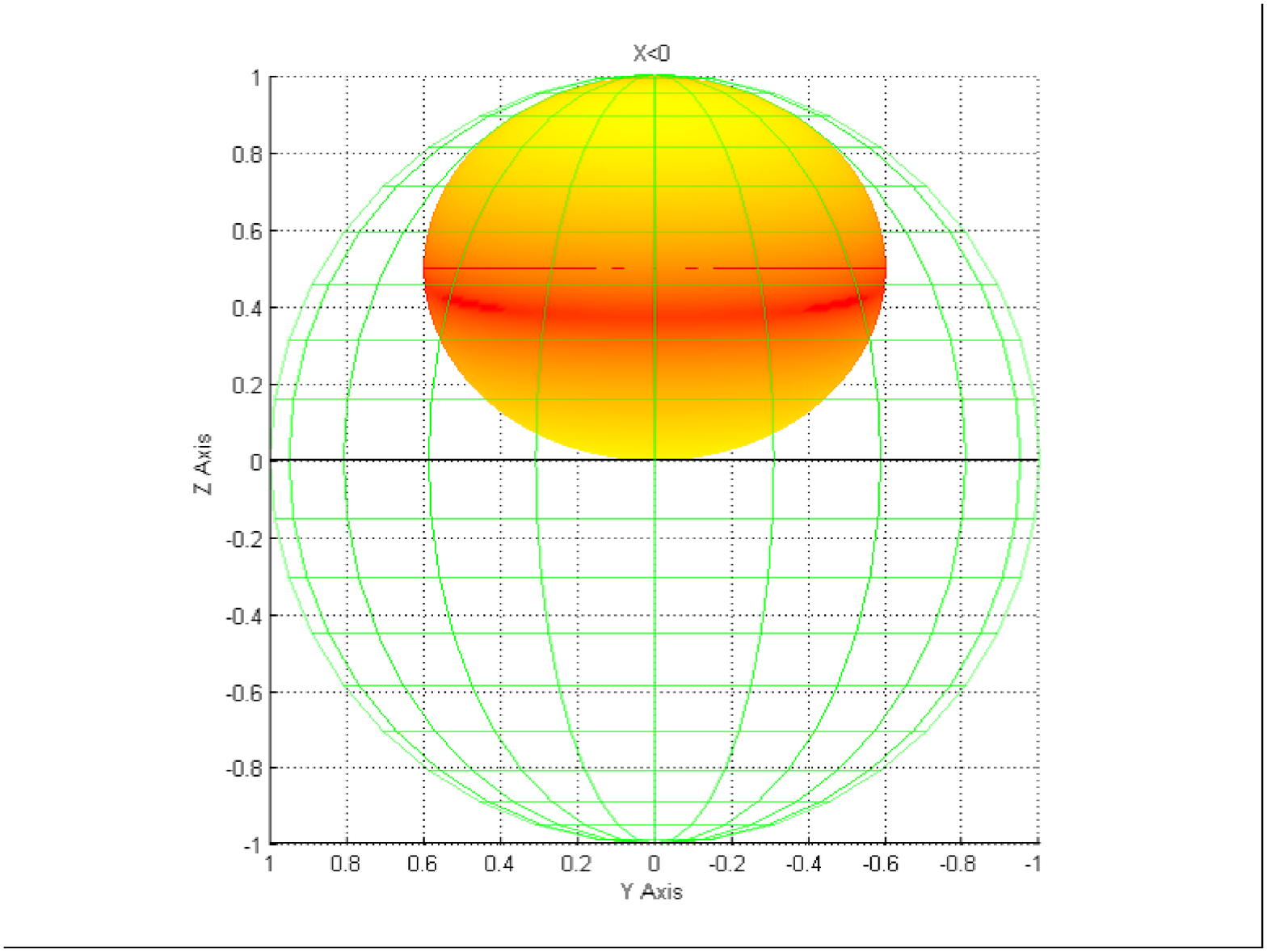}
 for   images of the two hemispheres of the Bloch sphere.    ~~left:  $x > 0$~~~~right:  $x < 0$ \\
  \end{center}
   \vskip0.5cm    
      Scale: 
 \vskip-0.9cm~~~~~~~~  \includegraphics[height=1.8cm,keepaspectratio=true]{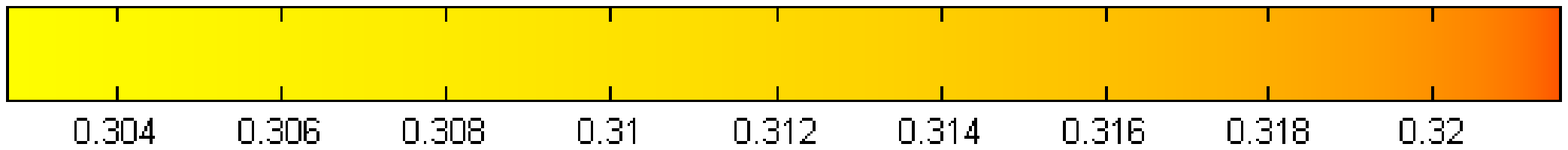}
   \caption{Relative entropy $H[\Gamma(\omega), \Gamma( \rho^3_{\av})] $
 with respect to the 3-state average output $ \Gamma( \rho^3_{\av})$
  for image states $\Gamma(\omega)$.   Note that this figure is almost indistinguishable
  from  Figure~\ref{fig:four}.   However, the actual locations
  and values are slightly different as seen by comparing the values in Table~\ref{tab:relent} below 
  with those in Table~\ref{tab:4st}  }
   \label{fig:relent2}.
  \end{figure}

The 4-state optimal ensemble satisfies
this requirement, and  $H[\Gamma(\rho_i), \Gamma( \rho_{\av}^4)] = 0.321485159 $
for all $i$. 
  If, instead, the 3-state ensemble for the same channel
  (i.e., the last reported in Table~\ref{tab:3st}) is used, one finds that
  $H[\Gamma(\rho_i^3), \Gamma( \rho_{\av}^3)] = 0.321460988 ~\forall ~i$
so that these  states also satisfy the equi-distance requirement.    However, as
one can see from Table~\ref{tab:relent}
  $$  \sup_{\omega}  H[\Gamma(\omega), \Gamma( \rho_{\av}^3)]  ~ > ~  0.3215
    ~ > ~     H[\Gamma(\rho_i), \Gamma( \rho_{\av}^3)]  $$
  showing that the 3-state ensemble is {\em not}  optimal.   Indeed, a plot of
  $H[\Gamma(\omega), \Gamma( \rho)]$ as shown in Figure~\ref{fig:relent2}, shows 
  four relative maxima, which lie closer to the 4-state inputs, than to the
  3-state inputs for which $y_i = 0$.   The supremum appears to be
  achieved for a pair of states  with $(x,y,z) =(-0.539291 , \pm 0.822613 , -0.180202) $.
  Thus, the relative
  entropy criterion seems to anticipate the splitting of the input near
 $ (-0.97,0,-0.25)$ into a pair of inputs near $(-0.47,\pm 0.86,-0.17)$.

\begin{table} [p]   \begin{center}
$$  \begin{array}{ccc} 
  \omega(x,y,z)  &  H[\Gamma(\omega), \Gamma( \rho_{\av}^3)] & H[\Gamma(\rho_i), \Gamma( \rho_{\av}^3)]  \\
(0.252867, 0.000000 , 0.967501 )  &0.321460988 &  0.321460986\\
(0.978544, 0.000000 , 0.206036 )  &0.321460988 &    0.321460981 \\
(-0.539291 , 0.822613 , -0.180202)   &0.321505535 &  0.321504592 \\
(-0.539291 ,- 0.822613 , -0.180202)   &0.321505535 &  0.321504592 \\
 \end{array} $$
 \caption{Relative maxima of relative entropy with respect to  the 3-state $\rho_{\av}^3$ for 4-state
 channel  $\Gamma(\rho(x,y,z))=\rho(0.6 x+ 0.021, 0.601 y, 0.5 z +0.495)$.   The relative entropy for the nearest 4-state input is also  given for comparison.}
 \label{tab:relent} \end{center}   \end{table}

 The relative entropy can also be used to check additivity without need
 to carry out the full variation in (\ref{eq:maxmin}).  In fact, applying
  (\ref{eq:relentsqz}) to the product channel $\Gamma \ot \Gamma$ gives
  \begin{eqnarray}  \label{eq:reletntest}
     2 C(\Gamma) \leq C(\Gamma \ot \Gamma) \leq \sup_{\omega}
        H\big[ (\Gamma \ot \Gamma)(\omega) \, , \, \Gamma(\rho_{\av}^4) \ot \Gamma(\rho_{\av}^4)\big].
  \end{eqnarray}
 If the supremum on the right equals $2 C(\Gamma) $, then the channel is 
 additive.  Furthermore,  the supremum restricted to product inputs
 equals $2 C(\Gamma) $.
 Therefore, if the supremum is strictly greater
 than $2 C(\Gamma) $, it must be attained for a pure entangled state
 $\omega$.     But this would imply that the optimal average input is
 not a product and, hence, that $\Gamma$ is superadditive.   Thus, to
 determine whether or not additivity holds,  it is enough to study the
 supremum in (\ref{eq:reletntest}) for the product input $\rho_{\av}^4 \ot \rho_{\av}^4$; it is
 {\em not} necessary to find the optimal inputs for the product channel.

In order to reformulate the relative entropy optimization in terms of a hyperplane 
condition, we introduce some notation and review some elementary facts.
First, recall that
${\rm Tr}\, A^{\dagger} B = \sum_{jk} \overline{a}_{jk} b_{jk} = {\bf  \overline{a}} \cdot {\bf b}$
where {\bf a, b} denote  vectors with components $a_{jk}$ and $b_{jk}$ respectively.
Alternatively, let $\{ M_k \}_{k = 0,1 \ldots d^2-1}$ be an  orthonormal basis
of $d \times d$  matrices with ${\rm Tr}\, M_j^{\dagger} M_k = \delta_{jk}$
and $M_0 = \frac{1}{d} I$.   Then an arbitrary matrix $A$ can be written as
$A = \sum_k \alpha_k M_k$ with $\alpha_k =  {\rm Tr}\, M_k^{\dagger} A$,
and ${\rm Tr}\, A^{\dagger} B = \sum_k    \overline{\alpha}_k \beta_k$.
 A familiar example of such a basis for $2 \times 2$ matrices is
  $\{ \half  \sigma_0, \half \sigma_1, \half \sigma_2, \half \sigma_3 \} $ where 
 $\sigma_0 $ denotes the identity $I$.    An example for $4 \times 4$
matrices is $\{ \frac{1}{4} \sigma_k \sigma_k \}_{j,k = 0,1,2,3}$
We will be primarily interested in basis and matrices which, like the two
examples above,  are self-adjoint;
therefore, we drop the adjoint symbol $~^\dagger$ and assume the
coefficients $\alpha_k $ are real. 
For a density matrix  $\rho$ we will let $\beta(\rho)$ be the vector associated with
the trace zero part of $\rho$ so that
$\rho = \frac{1}{d} I + \sum_k  \beta_k M_k$.
Using the Pauli basis for  qubits, $\beta(\rho)$
  is simply the vector with components $(x,y,z)$ in
${\bf R}^3$ associated with the Bloch sphere. 

Now let $F(\rho) =   S[\Gamma(\rho)] - \xi \cdot  \Gamma(\rho)$ with $\xi$ defining
a supporting
hyperplane for the capacity optimization as discusses in Section~\ref{sect:conv},
and  let $ G(\rho) = H[\Gamma(\rho), \Gamma(\rho_{\av})] $ with $\rho_{\av}$ the
optimal average.  Writing
$ \log \Gamma(\rho_{\av}))  = \sum_k \tau_k M_k$, one finds
\begin{eqnarray} \nonumber
G(\rho) =  H[\Gamma(\rho), \Gamma(\rho_{\av})]  & = & -  S[\Gamma(\rho)] - {\rm Tr} \, \Gamma(\rho)     \, \log(\Gamma(\rho_{\av}) ) \\
   & = &  -  S[\Gamma(\rho)] - \tau_0 - \tau \cdot \beta(\Gamma(\rho)).
\end{eqnarray}
Therefore, $H[\Gamma(\rho), \Gamma(\rho_{\av})]  +  S[\Gamma(\rho)] $ defines
a hyperplane and $G(\rho) + \tau_0 \leq C(\Gamma) $ holds with equality for the
optimal inputs $\rho_i$.   This implies that the supporting hyperplane
condition $F(\rho) \geq A$ holds with equality for optimal inputs $\rho_i$
 when $\xi = -\tau$.  In that case, $F(\rho) = -G(\rho) - \tau_0$
and $A = \tau_0 - C(\Gamma)$.   With $d+1$ optimal inputs, the supporting
hyperplane is the unique hyperplane given by the relative entropy.

For the 4-state channel $\xi = (0.039662, 0, 0.962107)$ and we
see from Table 1 that $A = 0.978506$ and $B = 0.321485$.   A computation gives
$\log(\Gamma(\rho_{\av}) ) = 1.299989 I +0.039662 \sigma_x + 0.962105 \sigma z$,
from which it follows immediately that  $\tau = - \xi$ and  $F(\rho) = 1.299989 - G(\rho)$
as expected.
 
\bigskip  
 \section{Additivity}

As mentioned earlier, 4-state channels might be good candidates for
examining the additivity of channel capacity.   Those considered here
have the property $\lambda_2 > \max_{i = 1,3} | \lambda_i | $, $t_2 = 0$
and $t_1, t_3 \neq 0$.    Channels of this type do not belong to one of
the  classes  of qubit maps for which multiplicativity of the maximal p-norm
has been proved and its geometry seems resistant to simple analysis.  (See
\cite{king-ruskai2} for a summary and further references.)   Because one
state lies very close the the Bloch sphere, with all others  much further
away, one expects that additivity of minimal entropy and multiplicativity 
of the maximal p-norm surely hold for this channel.   Nevertheless,
this has not been proven, suggesting
 that the channel may have  subtle properties.   Indeed, 
most known proofs of  additivity for minimal entropy 
for a particular class of channels, also yield additivity of channel capacity
for the same class.   These conjectures
are now known to be equivalent \cite{shor2}, but this equivalence 
requires the use of non-trivial channel extensions and does 
not hold for individual channels.       Thus the resistance to proof of
of a seemingly obvious fact using current techniques may 
indicate that the far less obvious additivity of channel capacity
does not hold.

\medskip

We will use the fact   that $\Gamma$ is additive if
$\sup _{\omega} G(\omega) = 2 C(\Gamma)$, but superadditive if  
$G(\omega) > 2 C(\Gamma)$ for some state $\omega$ where
 $G(\omega) = H\big[ (\Gamma \ot \Gamma)(\omega) \, , \, \Gamma(\rho_{\av}^4) \ot \Gamma(\rho_{\av}^4)\big].$
The function $g(\rho) = H[\Gamma(\rho), \Gamma(\rho_{\av}^4)]$ has 10 critical
points(4 maxima, 4 saddle points, and 2 (relative) minima), as shown   in Figure~\ref{fig:relent3}.
This implies that  $G(\omega)$ has at least 100 critical points, 
16 maxima, 4 (relative) minima, and 80 saddle-like critical points when one
restricts $\omega$ to a product state.   The complexity of this landscape
seems greater than that of any other class of channels studied.
If the capacity of any  qubit channel is non-additive, it seems 
likely that it would be a channel of this type.   Therefore,
 a thorough numerical analysis is called for.    Unfortunately,
the large number of critical points, also make a full optimization
very challenging.

\medskip

It suffices to optimize over  pure states of the form
$\omega = |\Psi \rangle \langle \Psi |$ with 
\begin{eqnarray}   \label{eq:4dimrep}  
   |\Psi \rangle = \sqrt{p}   \pmx  ~~ \cos \theta_u \\  e^{i \phi_u}  \sin \theta_u \emx \ot
       \pmx   ~~ \cos \theta_v \\ e^{i \phi_v}   \sin \theta_v \emx +  e^{i \nu}
  \sqrt{1-p}       \pmx    e^{-i \phi_u }   \sin \theta_u \\  ~~ -  \cos \theta_u \emx \ot
       \pmx    e^{-i \phi_v }  \sin \theta_v \\  ~~ -  \cos \theta_v \emx
\end{eqnarray}
and $p \in [0,1], ~ \theta_u, \, \theta_v, \, \nu  \in [0,2 \pi], ~ \phi_u, \, \phi_v \in [0,\frac{\pi}{2}]$.
To see why this is true, note that (\ref{eq:4dimrep}) says that
$|\Psi \rangle = \sqrt{p} \, |u \rangle \ot   |v \rangle + e^{i \nu} \sqrt{1-p}   \,
   |u^{\perp}  \rangle \ot   |v^{\perp}  \rangle $ where $ |u^{\perp}  \rangle $
   denotes the vector orthogonal to 
   $| u \rangle =  \pmx  ~~ \cos \theta  \\  e^{i \phi }  \sin \theta  \emx$.  Note that 
$$ |u \rangle \langle u | = \half \big[ I + \sin 2 \theta \, \cos \phi \, \sigma_x 
    + \sin 2 \theta \, \sin \phi \, \sigma_y + \cos 2 \theta \, \sigma_z \big]~.$$
 
Now let  $\gamma_u = \Gamma( |u \rangle \langle u|)$.  Then we can write
\begin{eqnarray}
 ( \Gamma \ot \Gamma )(| \Psi \rangle \langle  \Psi | )=
      p \, \gamma_u \ot \gamma_v + (1-p) \, \gamma_{u^{\perp}} \! \ot  \gamma_{v^{\perp}}  
         + \sqrt{p(1-p)} \,  X
\end{eqnarray}
where
$$ X = e^{-i \nu} \Gamma( |u \rangle \langle u^{\perp} |) \ot \Gamma( |v \rangle \langle v^{\perp} |) +
  e^{ i \nu} \Gamma( |u^{\perp} \rangle \langle u |) \ot \Gamma( |v^{\perp} \rangle \langle v |) $$
Since ${\rm Tr} \,| u \rangle \langle u^{\perp} | = \langle u^{\perp} | u \rangle = 0$ and
$\Gamma $ is trace-preserving, the partial traces of $X$ are zero, i.e.
 \begin{eqnarray}
      {\rm Tr} _1 \, X =   {\rm Tr} _2 \, X  = 0 ~.
 \end{eqnarray}
It then follows immediately that  ${\rm Tr}  \,  X  \, \log \varrho_1 \ot  \varrho_2  = 0$ since
  \begin{eqnarray}
         {\rm Tr}  \,  X  \,  I_1  \ot  \log  \varrho_2  +   {\rm Tr}  \,  X  \,  ( \log  \varrho_1 ) \ot I_2  
    =    {\rm Tr}_2  \big [ \log  \varrho_2  ( {\rm Tr} _1 \, X)  \big] +  
         {\rm Tr}_1   \,  \big[ \log  \varrho_1   ( {\rm Tr}_2 X)  \big] = 0
 \end{eqnarray}
Applying this with $\varrho =   \Gamma(\rho_{\av}^4)$ one finds  that 
  \begin{eqnarray}  \label{eq:trX0}
  {\rm Tr } \,   (\Gamma \ot \Gamma) (| \Psi \rangle \langle  \Psi |  ) \, \log \Gamma(\rho_{\av}^4) \ot \Gamma(\rho_{\av}^4)  = 0 
    \end{eqnarray}
  Therefore the  second term in the relative entropy is affine  in $p$.  Hence  
any non-linearity in  \linebreak
$H(\Gamma \ot \Gamma) (| \Psi \rangle \langle  \Psi |  ) ,  \Gamma(\rho_{\av}^4) \ot  \Gamma(\rho_{\av}^4) $ must come entirely from the entropy term
$-S\big[(\Gamma \ot \Gamma) (| \Psi \rangle \langle  \Psi |  )\big]$.

\begin{figure}[h]  \begin{center}
  \includegraphics[width=7.5cm]{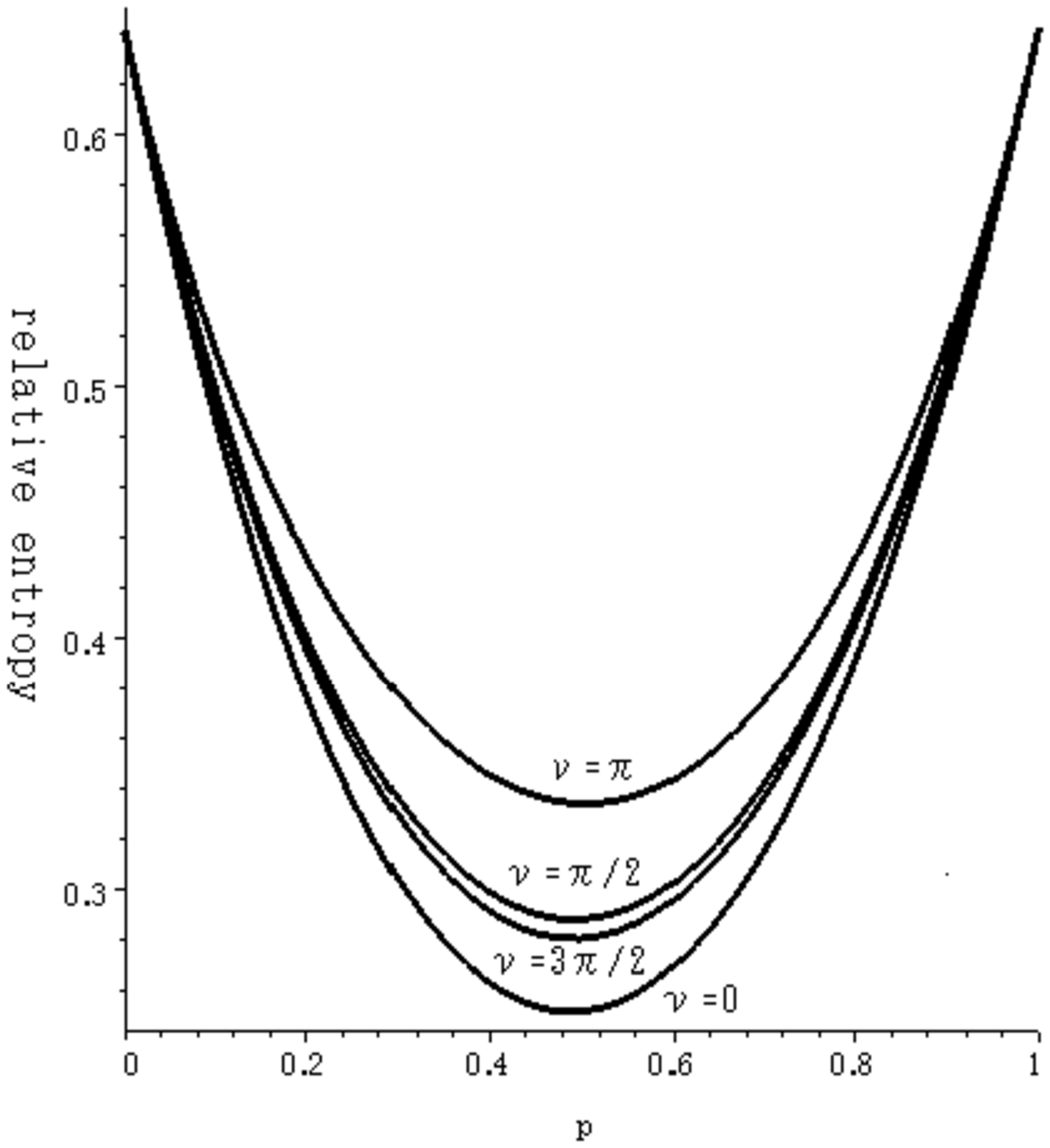} \end{center}
 \caption{Typical plot of  $G(\omega) =  
  H \big[ (\Phi \ot \Phi)(\omega) , \Phi(\rho_{\av}) \ot  \Phi(\rho_{\av}) \big]$
   as of function of $p$ for  $\nu = 0, \frac{\pi}{2}, \pi, \frac{3\pi}{2}$ using pure states of the form
 (\ref{eq:4dimrep}) 
   and $u, v$ fixed and $e^{i \nu} = 1, i, -i, -1$.   Endpoints correspond to product states and $p = 0.5$ 
   maximally entangled .}
   \label{fig:pcrelent}
  \end{figure}

Because of the difficulty of optimizing over all six parameters, plots of $G(\omega)$
were made as a function of only $p, \nu$ with $u,v$ fixed and as a function of $p$
with the remaining 5 parameters fixed.   A typical example is shown in Figure~\ref{fig:pcrelent} 
and appears to be  convex function in $p$ for several choices of $
nu$.    Many other examples were considered
with   $u ,v $ both corresponding to optimal inputs,
 $u ,v $ chosen randomly,
 $u ,v $ chosen to be highly non-optimal, and
 various combinations of these.
The shape of the curve seems to be extremely resilient for all inputs
in Schmidt form (\ref{eq:4dimrep}) and  
suggests convexity in $p$ with a deep minimum.
  Although the minimum lies above that for the corresponding
mixed state with $X = 0$, it is well below both endpoints.    
Changes as $\nu$  ranges from $0$ to $2 \pi$ are small.    

States of the
    form   $\frac{1}{\sqrt{2}} \big( | u_i \rangle \ot  | u_j\rangle + e^{ i \nu}| u_k \rangle \ot  | u_{\ell}\rangle \big)$
    with $u_i$ corresponding to the four optimal inputs were also considered.    
    Because these $u_i$ are
    not orthogonal, the functions do not have the form (\ref{eq:4dimrep}) and 
     (\ref {eq:trX0}) need not hold.   Although the 
     relative entropy   has a slightly different shape as a function of $p$ and $\nu$,
     it still lies below the plane $2 C(\Gamma)$ and has a deep minimum.

Thus, there seems to be little room  
 for obtaining a counter-example by varying the channel parameters.
This may give the strongest numerical evidence for additivity yet, at least
in the case of qubit channels.

\medskip

\noindent{\bf Remark:}  Because the second term in the relative entropy
is affine in $p$ for states of the form (\ref{eq:4dimrep}), the
concavity of the entropy function 
$g_{u,v, \nu} (p) \equiv S\big[(\Gamma \ot \Gamma) (| \Psi \rangle \langle  \Psi |  )\big]$
as a function of $p$ for arbitrary states of the form form (\ref{eq:4dimrep}).   
This would immediately yield both additivity of minimal entropy and of
channel capacity.     It is very  tempting to conjecture that $g_{u,v, \nu} (p)$
is concave.

\begin{figure}[h]     \vskip-0cm    \hskip4cm
  \includegraphics*[width=9cm,height=9cm,keepaspectratio=true]{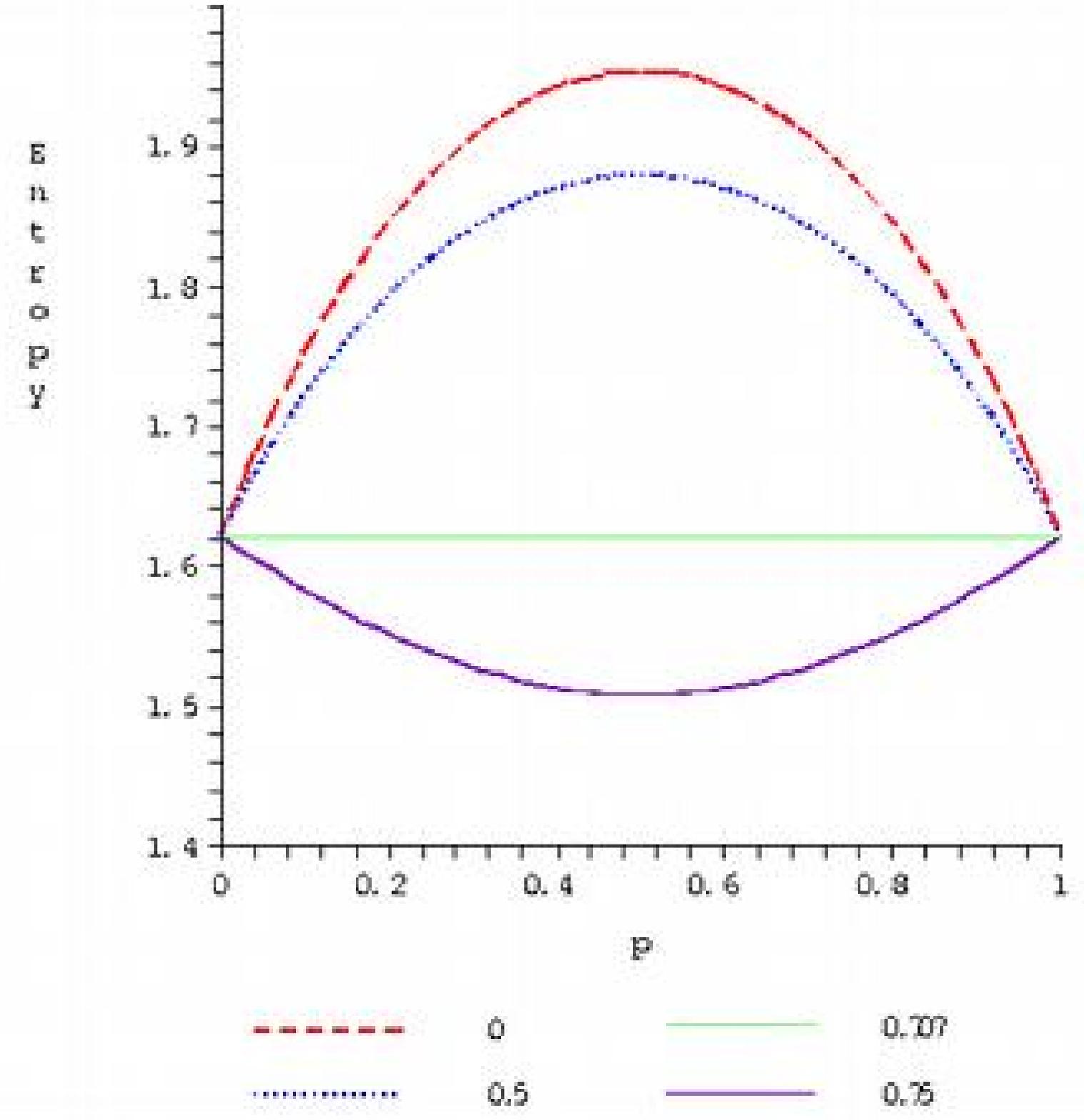} 
  
 \caption{Plots of $f(p) = 
     S \big[ (\Gamma \ot \Gamma) \big(  |\psi \rangle \langle \psi | \big) \big]$
 for $\mu = 0, 0.5, 0.707, 0.75 $ with 
 $\Gamma[\rho(x,y,z)] = \rho(\mu x,  \mu y, 0.5 x)$ and
 $|\psi \rangle = \sqrt{p} \, |00 \rangle  +  \sqrt{1-p}\,  |11 \rangle $.
 The top curve with $\mu = 0$ reduces to the usual concavity
 of the mixed state  $(\Gamma \ot \Gamma)( p |00 \rangle \langle 00| + (1-p) |11 \rangle \langle 11|$;
 the next with $\mu = 0.5$ shows the expected concavity;  
 the flat horizontal curve is for $\mu=0.707$, or $\sqrt{2}$;
 the bottom curve shows $\mu = 0.75$ for which the inputs are no longer optimal
 and   $f_{\mu}(p) $ is convex. }
  \label{fig:concex}
  \end{figure}

A similar conjecture  was made independently in \cite{DHS} with supporting
evidence for a particular  set of channels with $d > 3$.   Despite the appeal of this
conjecture, it is false.      Consider the channels 
 $\Gamma[\rho(x,y,z)] = \rho(\mu x,  \mu y, 0.5 x)$ with $0 \leq \mu \leq 0.75$
 and $|\psi \rangle = \sqrt{p} |00 \rangle  +  \sqrt{1-p} |11 \rangle $. Then
 $\Gamma \ot \Gamma) \big(  |\psi \rangle \langle \psi | \big)$ has eigenvalues
 $\frac{3}{16}, \frac{3}{16},  \frac{5 \pm 4 \sqrt{1 + (16 \mu^4 - 4)p(1-p)}}{16}$.  It 
 follows  that $f(p) = S \big[ (\Gamma \ot \Gamma) \big(  |\psi \rangle \langle \psi | \big) \big]$
 is concave for $\mu \leq \frac{1}{\sqrt{2}}$ and convex  for $\mu \geq \frac{1}{\sqrt{2}}$ as shown in Figure~\ref{fig:concex}.   This example above also implies that a related conjecture \cite{DHS} for  Schur concavity is false.    
  Note however, that the chosen inputs are not optimal when $\mu  > \frac{1}{2}$ and far
  from optimal when $\mu > \frac{1}{\sqrt{2}}$; 
  indeed even the lowest point on convex curve shown lies well above the true minimal output 
  entropy of 1.2017521 for $\mu=\frac{1}{\sqrt{2}}$ and 1.087129 for $\mu=0.75$.
  
       If products of the 
optimal inputs  $\frac{1}{\sqrt{2}} ( |0 \rangle  \pm  |1 \rangle)$ are entangled, the
 corresponding entropy function is known \cite{king-ruskai} to be concave.
 Moreover, King has \cite{king} shown  that both the minimal entropy and the capacity 
  are additive for these channels for all $\mu$.
  
 It seems likely that the conjectured concavity holds when optimal inputs are
 entangled; however, this is not sufficient to prove additivity of either capacity
 or minimal entropy.  
 
\subsection*{Acknowledgment}

The authors would like to thank Dr. Mitsuru Hamada for his useful
comments on this problem. This work was begun when M.B.R. visited the
ERATO in September, 2003, and some of the work of T.S. was done during 
a visit to Tufts University in February, 2004. 


\end{document}